\documentclass[twocolumn]{aastex631}

\usepackage{natbib}
\usepackage{graphicx}
\usepackage{graphics}
\usepackage{amsmath}
\usepackage{color}
\usepackage{hyperref}
\definecolor{lblue}{rgb}{0,0.2,0.6}
\definecolor{dgreen}{rgb}{0.1,0.6,0.3}
\usepackage{enumitem}

\usepackage{nameref}
\usepackage{varioref}
\usepackage{hyperref}
\usepackage{cleveref}
\usepackage{amsmath,amssymb,amsxtra,amsfonts}   
\usepackage{txfonts}

\newcommand       \be           {\begin{equation}}
\newcommand       \ee           {\end{equation}}

\begin{document}

\title{The faint end of the UV luminosity function at $0.4<\lowercase{z}<0.7$ from the Hubble Frontier Fields}

\correspondingauthor{Lei Sun, Xiao-Lei Meng, Hu Zhan}
\email{sunl@ucas.ac.cn \quad mengxiaolei@bao.ac.cn  \quad zhanhu@nao.cas.cn}

\author[0009-0004-6325-7839]{Lei Sun}
\affil{School of Astronomy and Space Science, University of Chinese Academy of Sciences, Beijing 100049, China}

\author[0009-0006-0596-9445]{Xiao-Lei Meng}
\affil{National Astronomical Observatories, Chinese Academy of Sciences, Beijing 100101, China}

\author[0000-0002-9373-3865]{Xin Wang}
\affil{School of Astronomy and Space Science, University of Chinese Academy of Sciences, Beijing 100049, China}
\affil{National Astronomical Observatories, Chinese Academy of Sciences, Beijing 100101, China}
\affil{Institute for Frontiers in Astronomy and Astrophysics, Beijing Normal University,  Beijing 102206, China}

\author[0000-0003-1718-6481]{Hu Zhan}
\affil{National Astronomical Observatories, Chinese Academy of Sciences, Beijing 100101, China}
\affil{The Kavli Institute for Astronomy and Astrophysics, Peking University, Beijing 100871, China}

\author[0000-0002-8630-6435]{Anahita Alavi}
\affiliation{IPAC, Mail Code 314-6, California Institute of Technology, 1200 E. California Blvd., Pasadena CA, 91125, USA}

\author[0000-0003-4570-3159]{Nicha Leethochawalit}
\affiliation{National Astronomical Research Institute of Thailand (NARIT), Mae Rim, Chiang Mai, 50180, Thailand}

\author[0000-0002-4935-9511]{Brian Siana}
\affiliation{Department of Physics and Astronomy, University of California, Riverside, Riverside, CA 92521, USA}

\author[0009-0004-7133-9375]{Hang Zhou}
\affil{School of Astronomy and Space Science, University of Chinese Academy of Sciences, Beijing 100049, China}

\author[0009-0007-6655-366X]{Shengzhe Wang}
\affil{National Astronomical Observatories, Chinese Academy of Sciences, Beijing 100101, China}
\affil{School of Astronomy and Space Science, University of Chinese Academy of Sciences, Beijing 100049, China}

\author[0009-0003-1463-4397]{Shamuhawu Hailanhazi}
\affil{School of Astronomy and Space Science, University of Chinese Academy of Sciences, Beijing 100049, China}

\begin{abstract}
By extending the Hubble Frontier Fields (HFF) observations to the F225W band using HST WFC3/UVIS, we measure the rest-frame UV luminosity function (LF) of galaxies at $0.4<z<0.7$, pushing into the low-luminosity galaxy regime. In this first paper of a series, we describe the HST Cycle-27 GO-15940 F225W observations and data reduction, and present a corresponding catalog for the Abell 2744 field, which is the most data-rich HFF cluster field. Combining deep Near-UV imaging and the high magnification from strong gravitational lensing of the foreground cluster, we identify 152 faint galaxies with $-19.5 < M_{UV} < -12.1$ at $0.4 < z < 0.7$ through hybrid photometric-spectroscopic redshift selection from the Abell 2744 F225W catalog. Using a sample defined by a 50\% completeness cut and applying the maximum likelihood estimation, we derive the best-fit Schechter parameters for the UV LF at $z \sim 0.55$ down to $M_\text{UV} < -13.5$ mag, including a faint-end slope of $\alpha = -1.324^{+0.072}_{-0.074}$. We incorporate a curvature parameter $\delta$ in parameter estimation to account for a possible turn-over at the faint end of the UV LF, leveraging the exceedingly low luminosities probed by our sample. Our results rule out a turn-over brighter than $M_{UV} = -15.5$ at the $3\sigma$ confidence level.

\end{abstract}


\section{Introduction} \label{sec:intro}

Ultraviolet (UV) emission, primarily produced by young, massive stars, provides a direct observational proxy for recent star formation in galaxies. The UV luminosity function (LF)--defined as the number density of galaxies per unit UV luminosity--therefore offers a fundamental diagnostic for constraining the formation and evolution of galaxies across cosmic time \citep{Rees77, White78, Benson03}. Moreover, tracing the redshift evolution of the UV LF yields critical insights into the global star formation rate density of the universe and helps evaluate the role of galaxies, particularly those of low luminosity, in driving cosmic re-ionization. 

Our understanding of galaxy evolution has been significantly advanced by a wealth of new rest-frame UV luminosity function measurements, which now probe galaxies extending up into the cosmic dawn ($z>15$) \citep[e.g.,][]{Khusanova20, Rojas20, Bowler20, Ito20, Adams20, Adams23a, Adams.2024, Zhang21, Bouwens21, Bouwens.2022c, Harikane22, Harikane23, Harikane24, Harikane.2025, Finkelstein22a, Finkelstein22b, Finkelstein23, Leethochawalit23, Leung23, Perez23, Varadaraj23, Donnan23, Donnan.2024, Adak.2024, Willott.2024, Robertson.2024, Bagley.2024, Mcleod.2024, Perez.2025, Whitler.2025}. Within this rapidly evolving field, low-luminosity galaxies have received increasing attention. Their growing prominence is motivated by key roles across cosmic time: they are predicted to be dominant contributors to the ionizing photon budget at $z \sim 6-10$ \citep[e.g.,][]{Yung20b, Bouwens.2022c} and serve as essential probes of stellar feedback and re-ionization related processes at intermediate redshifts \citep[e.g.,][]{Yung20a}. The latest JWST surveys have pushed observational frontiers even farther, delivering robust LF measurements out to $z \sim 12-16$. For instance, the JWST Advanced Deep Extragalactic Survey (JADES) reports a high number density of galaxies around $-18 \leq M_{\rm uv}\leq -17$ at these epochs \citep{Whitler.2025}, suggesting a larger population of faint galaxies (as well as bright ones, e.g., \citealt{Donnan.2024}) than predicted by extrapolations from lower redshifts $z\leq9$ \citep{Bouwens21}. 

At low redshifts ($z < 2$) progress in constraining the UV LF has been achieved through a synergy of dedicated space-based observatories and deep, ground-based optical/UV surveys. Space missions such as XMM-Newton, Swift, and Astrosat have enabled targeted measurements within specific redshift windows, for instance around $0.4 < z < 1.2$ \citep{Page21,Sharma.2022,Sharma.2024,Bhattacharya.2024,Bhattacharya.2025,Belles.2025,Page.2025}. In parallel, wide-field ground-based programs like the CFHT Large Area $U$-band Deep Survey (CLAUDS) and the Hyper Suprime-Cam Subaru Strategic Program (HSC-SSP) have been utilized to trace the UV LF over a broader redshift baseline, extending from $z=0.2$ to $z=3$ \citep{Moutard20}. More recently, the Ultraviolet Imaging of the Cosmic Assembly Near-infrared Deep Extragalactic Legacy Survey (UVCANDELS) has started to provide high-resolution ultraviolet imaging, offering improved precision for studies at $z \sim 1$ \citep{Sun.2024}. Despite these advances, accessing the very faintest luminosities ($M_{\rm UV} > -15$) at low redshift continues to present a substantial observational challenge.

Pushing the frontier of galaxy detection to the lowest luminosities, especially at significant redshifts, requires surpassing the flux limits of conventional surveys. This is made possible by exploiting the magnifying power of strong gravitational lensing, in which foreground galaxy clusters serve as gravitational lenses, thereby enhancing the detectability of faint background sources. The feasibility of this approach was first demonstrated by the landmark discovery of a gravitationally lensed arc in the cluster Abell 370 \citep{Soucail.1987}. The methodology has since evolved through systematic surveys of lensing clusters. While surveys like the Cluster Lensing And Supernova survey with Hubble (CLASH; \citealt{Postman.2012}) provided valuable wider-field insights, it is the exceptional depth of the Hubble Frontier Fields (HFF; \citealt{Lotz.2017}) program that has uniquely enabled the systematic study of the faintest lensed galaxies. Leveraging this unprecedented dataset, numerous studies have successfully constrained the faint end of the UV luminosity function at redshift $z\sim 1-9$ \citep{Atek.2014,Atek.2015,Livermore.2017,Alavi16,Bouwens.2022c}. Extending such analyses to intermediate redshifts, \citet{Alavi16} measured the UV LF and its evolution at $1<z<3$ down to $M_{UV}=-12.5$ using two HFF clusters and Abell 1689. \citet{Bouwens.2022c} reported a steady steepening of $\alpha$ from $-1.53\pm0.03$ at $z=2$ to $-2.28\pm0.10$ at $z=9$, suggesting an increasing abundance of faint galaxies at higher redshifts. These deeper constraints are critical for refining the faint-end slope $\alpha$ and its redshift evolution, which in turn provides key insights into the cosmic star formation history and helps infer the role of faint galaxies during the epoch of cosmic re-ionization. 

The HFF program has assembled an unparalleled multi-wavelength dataset spanning from 0.2 to 8 $\mu m$ by combining deep observations from HST, Keck, VLT, and Spitzer. This legacy was further enhanced by the recent HST-GO-15940 program (PI: Ribeiro), which secured WFC3/UVIS F225W imaging for all six HFF clusters.
The addition of the F225W filter--the shortest wavelength UV filter available on HST--extends HST's continuous wavelength coverage from 1.7 down to 0.2 microns, enabling precise redshift determination for star-forming galaxies down to $z\sim1$. These advancements provide opportunities to study low-luminosity galaxy populations during a crucial epoch of cosmic evolution and to probe the faint end and possible turn-over of the UV LF at $z \sim 1$.

In this first paper of a series, we present the HST-GO-15940 F225W observations and data reduction. By focusing on the Abell 2744 field, the most data-rich Frontier Fields cluster including the latest JWST-GLASS NIRISS and NIRSPEC spectra, we analyze the rest-frame UV LF at $0.4<z<0.7$, leveraging this unique dataset to provide robust constraints on the LF parameters, with particular emphasis on the faint end.
The structure of this paper is as follows. We first describe the HFFs F225W observations and data reduction in Section~\ref{sec:obs-redu}. The photometry and catalog creation is then presented in Section~\ref{sec:photometry}. We next perform sample selection from the catalog of HFF A2744 field in Section~\ref{sec:sample}, with the corresponding completeness analysis detailed in Section~\ref{sec:comp}. We outline the procedures for deriving the best-fit UV LFs and present the results in Section~\ref{sec:lf}. These results are  discussed in Section~\ref{sec:discusstion}. We summarize and conclude in Section~\ref{sec:sum}. Throughout this paper, we assume $\Omega_m=0.3$, $\Omega_\lambda = 0.7$ and $H_0=70$ km s$^{-1}$ Mpc$^{-1}$ and all magnitudes used are AB magnitudes \citep{Oke83}.

\section{Observations and Data Reduction} \label{sec:obs-redu}
\subsection{Observations}\label{sec:obs}

The HST-GO-15940 program (PI: Ribeiro) obtained 8-orbit deep images with the WFC3/UVIS F225W band, alongside images with the ACS/WFC F475W band in parallel mode, complementing the existing data in all six clusters and parallel fields observed within the HFF program. 

To match the depths of existing WFC3/UVIS observations in the F275W and F336W bands in all six HFF clusters, 8 orbits of a total exposure time of 22674 seconds were secured in the F225W band, reaching a depth of 27.8 AB mag ($5\sigma$ limit in a $0\arcsec.2$ radius aperture). With multiple exposures ($\geq6$) per field, most cosmic rays can be effectively removed from the final data. Each visit consisted of a standard four-point dither pattern, with the center offset by 4$.\arcsec13$ ($\pm0.\arcsec42, \pm2.\arcsec02$ in POSTARG) to cover the chip gap and eliminate (or smooth) artifacts on scales smaller than the offset. According to the Space Telescope Science Institute’s best practices, a background level of $12~e^{-}$/pixel on average is suggested to mitigate the effects of UVIS charge transfer efficiency (CTE) degradation \citep{Mackenty.2012}. Given that the targets were selected to have low zodiacal light, we expected a background level of $6~e^{-}$/pixel, which was below the recommended value. We therefore employ full-orbit integrations to accumulate more dark current per exposure and minimize the required post-flash flux. To achieve the desired level across the entire CCD, we added an additional post-flash flux of $7~e^{-}$/pixel per orbit.

Given that full orbit exposures were being taken with the WFC3/UVIS camera, the ACS remained stationary during the orbit. We dedicated the same 8 orbits (divided into two sub-exposures per orbit to mitigate cosmic ray effects) to the F475W filter.  Like the F225W band, the F475W band also spans a wavelength range inaccessible to JWST. In the HFF parallel fields, this band has not previously been observed with HST, making its complement to the existing dataset, albeit at a $\sim0.7$ mag shallower sensitivity. 

\begin{figure}
\centering
\includegraphics[width=0.45\textwidth]{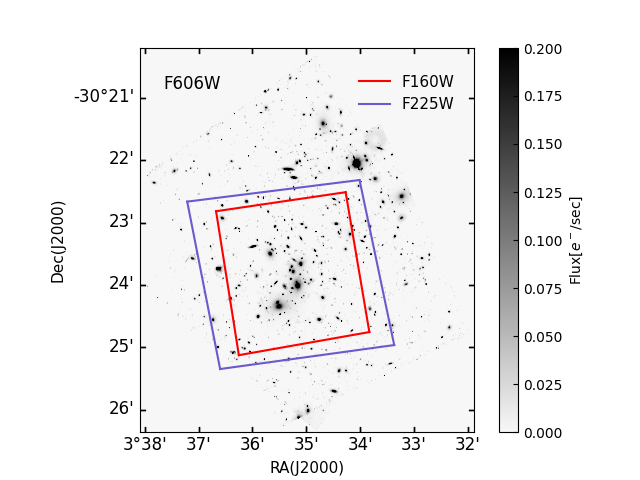}
\caption{Relative coverage of the F225W observations from HST-GO-15940 and the F160W observations in the Abell 2744 field, overlaid on the F606W mosaics. The F160W and F606 mosaics are provided by the HFF science products team \citep{Lotz.2017}.}
\label{fig:225}
\end{figure}

For the Abell 2744 field, which will be the focus of subsequent LF analyses, we summarize all the HST imaging observations available as of year 2025 and the corresponding depths in Table\ref{tab:obs} \citep{Alavi16, Lotz.2017}. In Figure~\ref{fig:225}, we also show the F225W footprints on top of the archival F606W mosaics of the Abell 2744 field.

\begin{deluxetable}{lll}
\tablecaption{Observations and Image Depths of the HFF Abell 2744 Field}
\tablehead{ 
\colhead{Instrument/Filter} & \colhead{Orbits} & \colhead{Depth\tablenotemark{a}} 
}
\startdata
WFC3/F225W & 8 & 27.8 \\
WFC3/F275W & 8 & 27.8 \\
WFC3/F336W & 8 & 28.2 \\
ACS/F435W & 18 & 28.7 \\
ACS/F606W & 9 & 28.7 \\
ACS/F814W & 41 & 29.0 \\
WFC3/F105W & 24.5 & 29.0 \\
WFC3/F125W & 12 & 28.6 \\
WFC3/F140W & 10 & 28.8 \\
WFC3/F160W & 24.5 & 28.8\\
\enddata
\tablenotetext{a}{$5\sigma$ limit in a $0.\arcsec2$ radius aperture.}
\label{tab:obs}
\end{deluxetable}

\subsection{Data Reduction}\label{sec:redu}

We performed the calibration of WFC3/UVIS and ACS/WFC data with customized routines based on the methods introduced by \citet{Rafelski.2015} and further developed by \citet{Prichard.2022}. To mitigate the progressive radiation damage affecting the WFC3/UVIS detector, we implemented the updated charge transfer efficiency (CTE) correction algorithm of \citet{Anderson.2021} on all F225W exposures. We also adopted the updated flux calibration tied to the most recent \texttt{CALSPEC} spectrophotometric models. This update is especially critical for ultraviolet measurements, as emphasized by \citep{2022AJ....164...32C}. Hot pixel masks were generated for each exposure using contemporaneous dark frames. The detection threshold was varied based on a pixel's distance from the readout amplifiers to ensure a consistent false-positive rate across the entire CCD\footnote{\url{https://github.com/lprichard/HST_FLC_corrections}}. A specific challenge in the F225W data is the presence of readout cosmic rays (ROCRs). These events occur when cosmic rays strike the detector after the amplifier readout sequence has initiated. The CTE correction algorithm mistakenly over-corrects for these hits, resulting in characteristic localized negative flux depressions, or “divots,” in the final image. We detected these ROCRs by searching for pixels that were $3\sigma$ negative outliers relative to the local background, located within a 5-pixel radius along the readout direction from primary cosmic-ray impacts already identified by the \texttt{AstroDrizzle} software \citep{Gonzaga.2012}. All pixels meeting these criteria were assigned a bad-data-quality flag in the final mask. Finally, we homogenized the background signal level across all four amplifiers of the UVIS detector. This step is crucial for producing final science images with a spatially uniform sky background\footnote{\url{https://github.com/bsunnquist/uvis-skydarks}}. 

For image registration and mosaicking, we employed a processing pipeline whose foundation was laid out by \citet{Alavi.2014}, with modifications tailored to our dataset. The core image combination was carried out with the \texttt{AstroDrizzle} software, which co-added the individually calibrated and flat-fielded exposures from both WFC3/UVIS and ACS/WFC. First, individual calibrated images within each visit were aligned to achieve relative astrometric alignment. To address the small offset in pointing and rotation between visits, \texttt{AstroDrizzle} was then run on each visit and the drizzled output image was aligned to the HFF astrometric reference grid. This alignment was performed on 30 mas/pix images with a precision of 0.15 pixels, utilizing unsaturated stars and compact sources. Finally, all aligned calibrated images were drizzled to the same pixel scale of 60 milliarcsec, consistent with the HFF reference images for the various fields. 

Alongside the science images, \texttt{AstroDrizzle} generates an inverse variance map, which is used to create weight images and calculate photometry uncertainties. Following \citet{Casertano.2000}, we implemented an additional correction to the weight images to account for the correlated noise. 

The HST data utilized in this study are available through the Mikulski Archive for Space Telescopes (MAST) at the Space Telescope Science Institute and can be accessed directly via \dataset[10.17909/t3pm-gy14]{https://doi.org/10.17909/t3pm-gy14}.

\section{Photometry and catalog creation} \label{sec:photometry}

We present our aperture-matched point spread function (PSF)–corrected photometry methodology used to produce photometric catalogs, focusing on the F225W catalog of the Abell 2744 field, which will be utilized for the LF measurements in subsequent sections. 

Recent studies conducted under the Ultra-Violet Ultra Deep Field (UVUDF; \citealt{Teplitz.2013}) and UVCANDELS projects have demonstrated that the UV-optimized aperture photometry method represents a more suitable approach for UV photometry compared to alternative techniques \citep{Teplitz.2013, Rafelski.2015, Sun.2024}. When running \texttt{SExtractor} \citep{Bertin.1996} in dual-image mode, this method employs the F606W band (V-band) as the detection image to extract object isophotes, rather than the generally-used F160W band (H-band). These isophotes are more compact than those typically obtained using the F160W and hence more appropriate for counting UV photons. Specifically in this work, we run \texttt{SExtractor v2.28.0} using the BCGs-out F606W mosaics that have been produced by the HFF science products team\footnote{\url{https://archive.stsci.edu/prepds/frontier/}} as the detection image. These mosaics have been processed to mask out bright central cluster galaxies (BCGs). For the measurement images, we use the PSF-matched F225W, produced from the original science images, convolved with PSF homogenization kernels, to bring their PSF's full width half maximum (FWHM) to match that of the F606W PSFs. We extract the F225W PSF by stacking all the unsaturated stars in the field, resulting in a FWHM of $0\arcsec.08$. The FWHM of the F606W PSF is measured to be $0\arcsec.112$ \citep{Pagul.2021}. The specific \texttt{SExtractor} parameter configuration employed for our F225W photometric analysis is presented in Table~\ref{tab:SE_params}. We refer to \cite{Sun.2024} for a detailed description of the photometry procedures.

Additionally, the F606W observations in the Abell 2744 field spans a larger overlapping area of 7.4 arcmin$^2$ with the F225W, compared to the F160W, which covers only 5.3 arcmin$^2$ within the F225W footprints, as illustrated in Figure~\ref{fig:225}. This further underscores the necessity of using the F606W as the detection band.

The flux uncertainties reported by \texttt{SExtractor} were derived from the RMS map that incorporates a correction for the pixel correlation introduced during the drizzling process (the \texttt{rms\_cor2} map), as described in Section~\ref{sec:redu}. The background noise standard deviation directly measured from our deep F225W science image (exposure time: 22,674~s) is $24.6~e^{-}$, which agrees well with the median noise level of $23.2~e^{-}$ from the RMS map. This corresponds to a mean background of $\sim 500~e^{-}$ under Poisson statistics. At this high mean level, the Gaussian noise approximation holds, thereby validating \texttt{SExtractor}'s error estimation for our data.

An aperture correction is then applied to get the F225W total magnitudes. We estimate magnitude uncertainties through proper error propagation. In total, we detect 1045 registered HFF sources in the Abell 2744 field, with a signal-to-noise ratio ($S/N$) threshold of $S/N\geq3$ in the F225W band.

\begin{deluxetable}{ll}
\label{tab:SE_params}
\tablecaption{\texttt{SExtractor} Parameter Configuration for WFC3/UVIS F225W Photometric Analysis.}
\tablehead{ 
\colhead{Parameter} & \colhead{Value}
}
\startdata
    \texttt{DETECT\_MINAREA}    & 9.0   \\
    \texttt{DETECT\_THRESH}     & 1.0 \\
    \texttt{ANALYSIS\_THRESH}   & 1.0  \\
    \texttt{FILTER\_NAME}       & gauss\_4.0 \\
    \texttt{DEBLEND\_NTHRESH }  & 64 \\
    \texttt{DEBLEND\_MINCONT}   & 1e-5 \\
    \texttt{BACK\_SIZE}         & 64 \\
    \texttt{BACK\_FILTERSIZE}   & 3 \\
    \texttt{BACKPHOTO\_THICK}   & 24 \\
\enddata
\end{deluxetable}

\section{Sample selection} \label{sec:sample}

The rich data sets in the HFF legacy, spanning across a wide wavelength range of $0.275-8\ \mu m$, enable robust photometric redshift (phot-z) estimates. \cite{Pagul.2021} derived the phot-zs for objects in all six cluster fields and their parallel fields. In comparison with the available spectroscopic redshifts (spec-zs) for the HFF clusters, this yields a combined outlier fraction of $10.3\%$ (defined as $|z_{\rm phot}-z_{\rm spec}| > 0.15 (1+z_{\rm spec})$) and a normalized median absolute deviation $\sigma_{\rm NMAD}=0.012$ after excluding the outliers. The GLASS-JWST ERS Program has significantly enriched the Abell 2744 field with recently released NIRSpec and NIRISS spectroscopic catalogs \citep{Mascia.2024,Watson.2025}, solidifying its position as the most extensively observed Frontier Fields cluster. Leveraging this unprecedented multi-wavelength dataset, we construct a hybrid-selected sample combining spectroscopic redshifts and photometric redshifts.

We adopt the absolute magnitude at the rest-frame wavelength of 1500\AA\ as our UV indicator, selecting sources whose 1500\AA\ flux is directly captured by the F225W filter to minimize K-corrections. This criterion defines our target redshift range as $0.4<z<0.7$, which is determined by the FWHM of the F225W filter. The Abell 2744 cluster itself has a redshift of $z=0.308$. All objects in our sample lie behind the cluster and can thus be properly treated as background sources in the lens-source system. 

\begin{figure}
\centering
\includegraphics[width=0.45\textwidth]{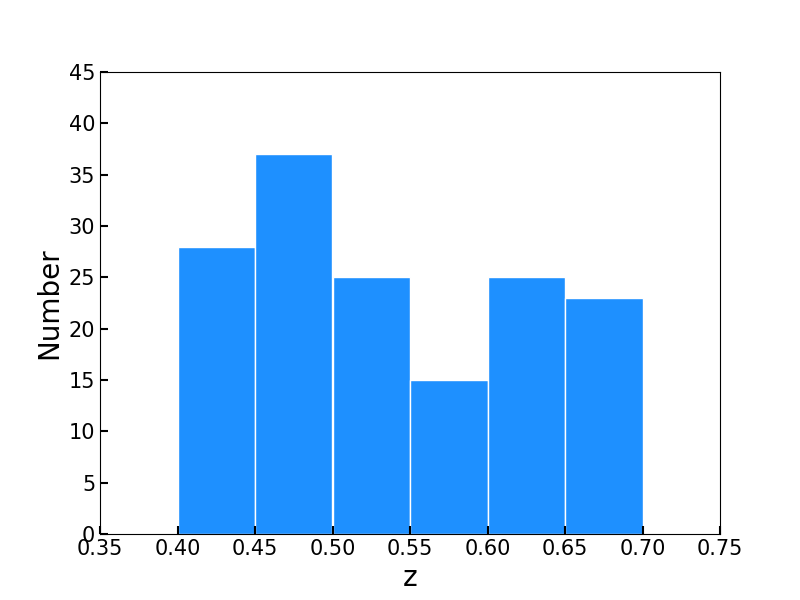}
\caption{Redshift distribution of the selected 152 sources from the Abell 2744 photometry catalogs, with $0.4<z<0.7$ and $S/N\geq3$ in the F225W band.}
\label{fig:Zdist}
\end{figure}

We first cross match our sources with the GLASS-JWST ERS NIRSpec and NIRISS spec-z catalogs, as well as a spec-z sample compiled from previous spec-z observations within the Abell 2744 field \citep{Owers.2011, Richard.2014, Mahler.2018, Wang.2020}. For sources with matched spec-z in the range $0.4<z_{\rm spec}<0.7$, we adopt their spec-z values. For sources lacking spec-z matches, we apply phot-z selection to use the same bounds $0.4 < z_{\rm phot} < 0.7$. We then visually inspect images from F225W (the measurement band), F606W (the detection band), and two additional bands with wavelengths shorter and longer than F606W—namely F435W and F814W—along with the corresponding F606W segmentation maps for all selected sources to reject spurious detections arising from segmentation failures. Our final sample consists of 152 sources within the target redshift range, including 34 with spectroscopically confirmed redshifts. Figure~\ref{fig:Zdist} shows the redshift distribution of these selected galaxies, which is a relatively flat distribution with a mean redshift $\sim0.55$.

In order to estimate the intrinsic luminosity of the background lensed galaxies in our samples, we require an accurate mass model of the galaxy cluster to calculate the lensing magnification. The HFF clusters have several independent lens models available, including Brada{\v{c}} models \citep{Bradac.2005, Bradac.2009,Hoag.2016n}, Natarajan \& Kneib (CATS) models \citep{Jullo.2009, Jauzac.2012, Richard.2014, Jauzac.2014, Jauzac.2015a, Jauzac.2015b, Lagattuta.2017}, Merten \& Zitrin models \citep{Merten.2009, Merten.2011, Zitrin.2009, Zitrin.2013}, Sharon/Johnson models \citep{Jullo.2007, Johnson.2014}, Williams models \citep{Liesenborgs.2006, Mohammed.2014, Jauzac.2014, Grillo.2015}, GLAFIC \citep{Oguri.2010, Ishigaki.2015, Kawamata.2016, Kawamata.2018}, Bernstein \& Diego models \citep{Diego.2005a, Diego.2005b, Diego.2007, Diego.2015}, keeton models \citep{Keeton.2010, Ammons.2014, McCully.2014} and Caminha models \citep{Caminha.2017}. We employ the latest V4C version of the Sharon/Johnson models. The magnification $\mu$ of sources can be calculated by
\begin{equation}
\mu = \frac{1}{|(1-\kappa)^2 - \gamma^2)|}\ ,
\label{eq:mu}
\end{equation}
where convergence $\kappa$ and shear $\gamma$ are derived from the normalized (to $z_{LS}/z_S =1$) $\kappa$ and $\gamma$ maps provided by the Sharon/Johnson V4C models.
\cite{Bouwens.2017} find that large systematic errors can occur at high magnifications ($\mu >30$) because of differences between the models. Since the highly magnified region around central bright galaxies is masked out, our final sample primarily consists of sources with low magnification $\mu < 7$, except for one source with $\mu$ in the range $[10,30]$. We therefore simply ignore this effect.

The rest-frame absolute magnitude at 1500\AA\ of the selected galaxy is then determined by
\begin{equation}
M_{1500}=m_{\mathrm{F225W}}+\mu_{\text{mag}}-5\text{log}(d_{L}/10\text{pc})-K_{\text{cor}}
\label{eq:M1500}
\end{equation}
\noindent where $\mu_{\text{mag}}$ is the magnification in units of magnitude, estimated using the lens models mentioned above. $K_{\text{cor}}$ represents the K-correction and is estimated for each individual galaxy using Equation 13 in \citet{Hogg02} based on their spectral energy distribution (SED) fitting results from the HFF archive \citep{Pagul.2021}. 

\begin{figure}
\centering
\includegraphics[width=0.45\textwidth]{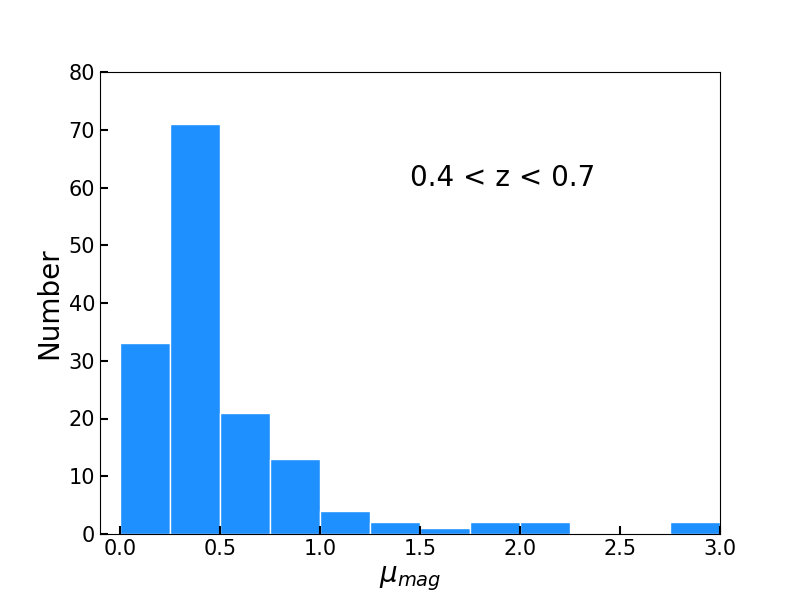}
\caption{Magnification (in units of magnitude) distribution of the selected 152 sources from the Abell 2744 photometry catalogs, with $0.4<z<0.7$ and $S/N\geq3$ in the F225W band.}
\label{fig:Mu}
\end{figure}

\begin{figure}
\centering
\includegraphics[width=0.45\textwidth]{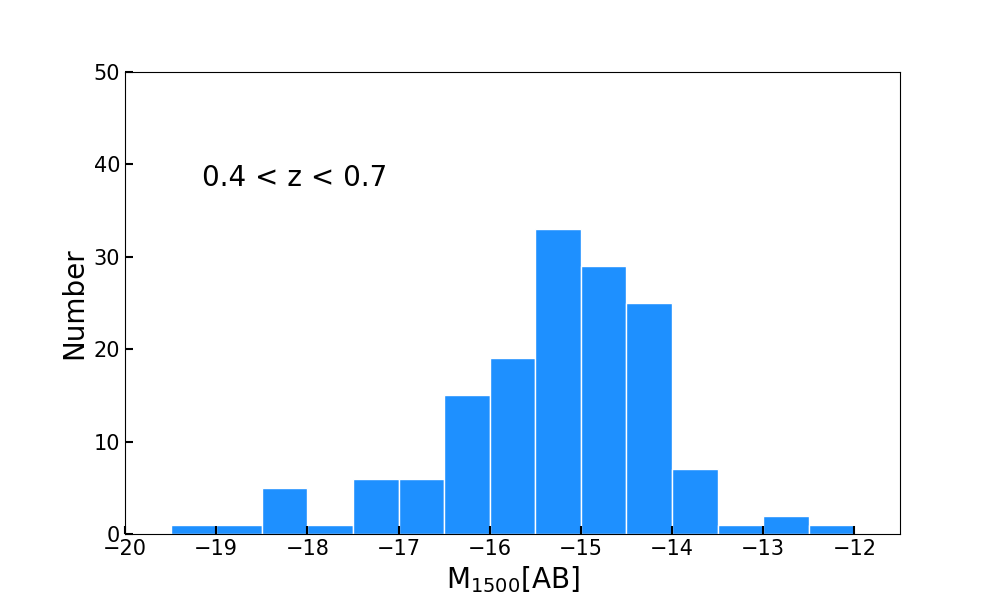}
\caption{Distribution of intrinsic absolute UV magnitude (corrected for the lensing magnification effect) of the selected 152 sources from the Abell 2744 photometry catalogs, with $0.4<z<0.7$ and $S/N\geq3$ in the F225W band. The faintest magnitude reaches $-12.1$.}
\label{fig:M1500}
\end{figure}

Figure~\ref{fig:Mu} shows the distribution of magnification (in units of magnitude) of galaxies at $0.4<z<0.7$ in the Abell 2744 sample. The magnitudes of magnification range between $0.5$ to $3$. After corrected for the lensing magnification effect, the distribution of intrinsic UV magnitude at rest-frame $1500 \AA$ of our sample is given in Figure~\ref{fig:M1500}, with a faint-end magnitude limit down to $-12.1$.

\section{Completeness corrections} \label{sec:comp}

To derive an accurate UV LF, which characterizes the underlying population of all star-forming galaxies within the survey volume, it is necessary to precisely estimate the completeness of a sample and apply corresponding corrections. At the low luminosity end of the LF, these corrections are particularly critical. A widely adopted method to estimate the completeness involves injecting artificial galaxies that resembles the underlying population into science images and calculate the fraction of recovered mock galaxies as a function of magnitude, redshift, etc \citep[e.g., ][]{Oesch10,Alavi16}, following the same data reduction and sample selection procedures as applied to the actual data. This technique is also applicable to studies in galaxy cluster fields involving the gravitational lensing effects, while properly accounting for the complexities arising from strong lensing magnification.
In this section, we perform a set of simulations to estimate the completeness in a 3-D grid of redshift $z$, intrinsic UV absolute magnitude $M_{\rm UV}$ and lensing magnification $\mu$.

First, building on \texttt{GLACiAR2} (\citealt{Leethochawalit.2022n}; see also its predecessor \texttt{GLACiAR}, \citealt{Carrasco.2018}), a publicly available pipeline for injection-recovery completeness simulations, we develop \texttt{GLACiAR2+}, an enhanced version applicable to the gravitationally lensed fields. We incorporate gravitational lensing effects by amplifying the flux and enlarging the size of the galaxies while maintaining constant surface brightness. For each source redshift $z_S$, we first compute the corresponding $\kappa_{z_S}$ and $\gamma_{z_S}$ maps by scaling the normalized (to $z_{LS}/z_S =1$) $\kappa$ and $\gamma$ maps from the Sharon/Johnson V4C lens models. 
For a mock galaxy at $z_S$, we then:
\begin{enumerate}[itemsep=0mm,label=\arabic*.]
\item Randomly pick a position $(x,y)$ in the image plane (masking out regions containing BCGs)      
\item Extract the corresponding $\kappa_{z_S}(x,y)$ and $\gamma_{z_S}(x,y)$ values    
\item Calculate the magnification $\mu$ using Equation~\ref{eq:mu}     
\item Apply the magnification by:    
\begin{itemize}[itemsep=0pt,topsep=0pt]
    \item Scaling the major and minor axes of the galaxy profile by $\sqrt{\mu}$
    \item Increasing the flux by a factor of $\mu$.
\end{itemize}
\end{enumerate}
Gravitational lensing induces anisotropic magnifications, distorting source shapes through tangential ($\mu_t$) and radial magnification ($\mu_r$) components \citep{Bartelmann.2010}. In our sample, the characteristic magnification of most our sample is small ($\mu=1-2$), the difference between $\mu_t$ and $\mu_r$ becomes negligible in our completeness simulations. We therefore adopt a single average magnification factor $\sqrt{\mu}$ for both major and minor axes.

We have also updated the intrinsic size distribution of source galaxies in \texttt{GLACiAR2+}. The completeness correction at faint magnitudes is critically sensitive to the assumed size distribution in the simulations. Adopting overly compact (extended) size distributions systematically biases the completeness estimates high (low), as demonstrated in \citep[e.g.,][]{Grazian.2011,Bouwens.2022a}. \citet[][hereafter Shibuya15]{Shibuya.2015} characterized the size distribution of galaxies across $0<z<8$ using the 3D-HST and CANDELS data, finding that star-forming galaxies follow a log-normal distribution in circularized effective radius ($r_e$) with the median decreasing toward high redshifts (at a given luminosity) and scaling with luminosity as $r_e \propto (L_{UV})^\alpha$ with $\alpha = 0.27$ across all redshifts. \cite{Bouwens.2022} demonstrated that at high redshift, faint galaxies ($M_{\rm UV}>-15$) are more compact than predicted by extrapolating the Shibuya15 size-luminosity relation, with $\alpha = 0.54$ at $z\sim4$ and $0.4$ at $z\sim 6-8$. Nevertheless, since our target sample lies at much lower redshift ($0.4 <z < 0.7$), we retain the Shibuya15 extrapolation in our completeness simulations. Specifically, we generate galaxy sizes at each luminosity and redshift bin by randomly sampling from the corresponding Shibuya15 log-normal distributions (Table 8 therein), extrapolating their relations for galaxies fainter than $M_{UV} < -16$. Using the randomly selected effective radius $r_e$, we then assign a S\'{e}rsic profile with index $n = 1.5$ for each galaxy, consistent with the Shibuya15 findings for star-forming galaxies.

Each simulated source galaxy is assigned a spectrum based on its redshift and absolute magnitude, randomly drawn from the JADES Extragalactic Ultra-deep Artificial Realization (JAGUAR) v1.2 catalog of mock star-forming galaxy spectra \citep{Williams18}. The integration of JAGUAR spectra constitutes a new feature in the recently updated \texttt{GLACiAR2} pipeline (Leethochawalit et al., in preparation). The observed flux in a certain band of a mock galaxy with intrinsic UV absolute magnitude $M_{\rm UV}$ is then obtained from the dot product of the assigned spectrum with amplitude increased by a factor of $\mu$ and throughput of this band. We insert stamps of artificial galaxies into both the F225W (for measurement) and the F606W (for detection) science images, after applying magnification factors to both size and flux and convolving the modified S\'{e}rsic profiles with band-specific point spread functions (PSFs). The simulated images are finally processed through the identical source identification pipeline (here \texttt{SExtractor}) as the original science images to generate catalogs. We derive the completeness function by computing the ratio of the number of detected mock galaxies to the number of injected. The detailed procedures can be found in \citealt{Carrasco.2018}.

\begin{figure}
\centering
\includegraphics[width=0.45\textwidth]{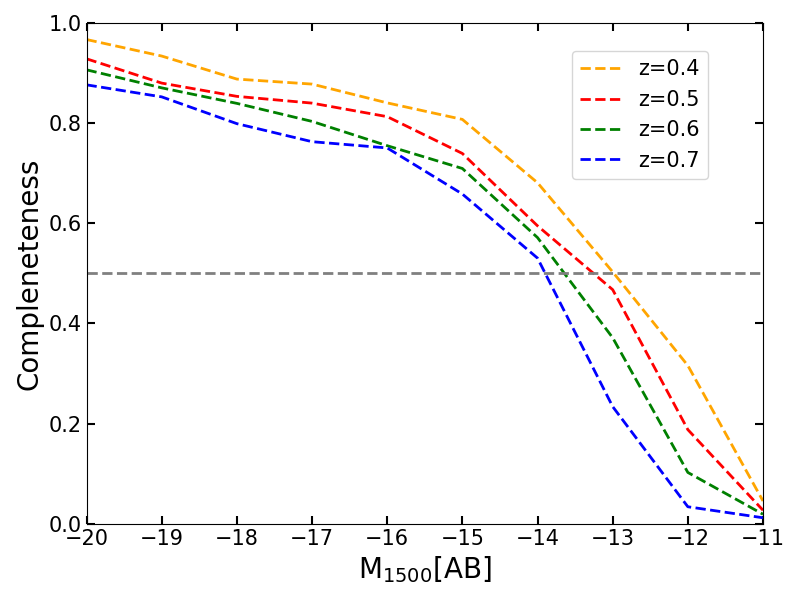}
\caption{Completeness as a function of absolute magnitude $M_{1500}$, with color-coded curves representing different redshifts. Here the magnification is marginalized over. The horizontal gray dashed line indicates our $50\%$ completeness threshold.}
\label{fig:C_M}
\end{figure}

Figure~\ref{fig:C_M} shows the calculated completeness functions from our simulations, where we present the completeness of different redshifts as a function of absolute magnitude (the color-coded curves), with the magnification marginalized over. The horizontal gray dashed line indicates our $50\%$ completeness threshold. To avoid using sources that require such large completeness corrections that the results become highly sensitive to uncertainties in the completeness estimation itself, we employ this completeness cut of $C > 50\%$ in the following analysis.

Figure~\ref{fig:C_M} demonstrates that the completeness at the bright end reaches $\gtrsim90\%$ when using F606W as the detection band. We emphasize that our completeness simulation, conducted by injecting mock galaxies into the science image of each band at random positions, naturally accounts for blending effects with foreground objects in the dense cluster field. As brighter galaxies generally have larger angular sizes, they are more susceptible to such blending, which can lead to detection rates below $100\%$. This effect is further amplified by the proximity of our target redshift range ($z\in[0.4,0.7]$) to the cluster's central redshift ($z=0.308$). In comparison, simulations using F160W as the detection image yield a significantly lower completeness of $\lesssim80\%$ even at the bright end, after excluding the effect of its smaller overlap with the F225W coverage (see Figure~\ref{fig:225}). This discrepancy likely stems from the larger isophotes typically extracted from the F160W image, which intensify blending with foreground cluster members. This further demonstrates the effectiveness of the F606W-based detection within our UV-optimized photometry approach.

To account for detection incompleteness, we compute the effective survey volume $V_{\mathrm{eff}}$ using the following relation, which incorporates the necessary corrections:
\begin{equation}
V_{\mathrm{eff}}(M)=\int_{0}^{\infty} \int_{z_{\mathrm{min}}}^{z_{\mathrm{max}}} \frac{d^{2}V_{\mathrm{com}}}{dz\ d\Omega}\ C(M,z,\mu)\ \Omega(z,\mu) \ \mathrm{d}\mu \ \mathrm{d}z\ ,
\label{eq:v_eff}
\end{equation}
where the integration limits $z_{\mathrm{min}}$ and $z_{\mathrm{max}}$ correspond to the redshift range spanned by our galaxy sample. The term $ \frac{d^2V_{\mathrm{com}}}{dz\, d\Omega} $ represents the comoving volume element per unit redshift interval per unit solid angle at a given redshift $z$. The completeness function $C(M,z,\mu)$ describes the probability of detecting a galaxy as a function of its absolute UV magnitude $M$, redshift $z$, and the gravitational magnification factor $\mu$. The function $\Omega(z, \mu)$ represents the differential area in the source plane (at redshift $z$) as a function of magnification $\mu$, i.e., $\Omega(z, \mu)\, d\mu$ is the area magnified by a factor between $\mu$ and $\mu+d\mu$. We derive this function for the Abell 2744 field from the magnification maps produced by the Sharon/Johnson V4C lens model, which provide the necessary mapping between image-plane positions and source-plane properties.

\section{Luminosity Function} \label{sec:lf}

A standard form used to model the galaxy luminosity function is the Schechter function \citep{Schechter76}, given by:
\begin{equation}
\phi(M)=0.4\ \mathrm{ln}(10)\ \phi^{*}\ 10^{-0.4(M-M^{*})(1+\alpha)}\ e^{-10^{-0.4(M-M^{*})}}\ .
\label{eq:sch}
\end{equation}
In this parameterization, $\alpha$ is the power-law slope at the faint end, $M^{*}$ denotes the characteristic magnitude and $\phi^{*}$ is the normalization factor.

To determine the Schechter parameters, we employ an unbinned maximum likelihood estimate (MLE) following \cite{Alavi16} and \cite{Sun.2024}. This method maximizes the joint likelihood function $\mathcal{L}$ for our sample, constructed as the product $\prod_{i=1}^{N} P(M_i)$ of individual detection probabilities, where the index $i$ runs over all $N$ galaxies brighter than our adopted magnitude limit. For a galaxy with observed absolute magnitude $M_i$, the probability $P(M_i)$ is defined as
\begin{equation}
    P(M_{i})=\frac{\int_{-\infty}^{+\infty} \phi(M) \ V_{\mathrm{eff}}(M)\ G(M|M_{i},\sigma_{i}) \ \mathrm{d} M}{\int_{-\infty}^{M_{\mathrm{limit}}} \phi(M) \ V_{\mathrm{eff}}(M)\ \mathrm{d} M}\ .
\label{eq:pm_mod}
\end{equation}
In this expression, the luminosity function $\phi(M)$ is taken from Equation~\ref{eq:sch}, while the effective volume $V_{\mathrm{eff}}(M)$ corrects for survey completeness. The upper integration limit $M_{\mathrm{limit}} = -13.5$ marks the faintest absolute magnitude in our sample. Finally, observational errors are included via the Gaussian distribution $G(M|M_i, \sigma_i)$, which models the uncertainty in each galaxy's magnitude:
\begin{equation}
    G(M|M_{i},\sigma_{i})=\frac{1}{\sqrt{2\pi}\sigma_{i}}\text{exp}(-\frac{(M-M_{i})^{2}}{2\sigma_{i}^{2}})\ .
\end{equation}

The absolute magnitude uncertainty $\sigma_i$ for each galaxy is the quadrature sum of three components: the photometric error $\sigma_m$, the redshift estimation error $\sigma_z$, and the uncertainty originated from lens models, $\sigma_{\mathrm{model}}$. The photometric term $\sigma_m$ is derived from the flux errors reported by \texttt{SExtractor}. The uncertainty in each galaxy's absolute magnitude due to photometric redshift errors is calculated by propagating the $1\sigma$ confidence interval of its redshift probability distribution through Equation~\ref{eq:M1500}. This uncertainty influences both the distance modulus and the magnification of each galaxy. To estimate the uncertainty from lens models, We employ 100 lower-resolution convergence and shear maps each corresponding to a model selected at equal intervals from the Markov Chain Monte Carlo (MCMC) chain, provided by the Sharon/Johnson V4C model.

As discussed in Section~\ref{sec:comp}, we apply a completeness cut of $>50\%$ to our sample to exclude objects requiring excessively large completeness corrections. This yields a final sample of 147 galaxies for deriving the UV LF , with a faint-end magnitude limit of $M_\mathrm{limit}=-13.5$. Both the sample size and magnitude limit are provided in Table~\ref{tab:pars}. Due to the limited survey area in the cluster field, which is further reduced when de-lensed back to the source plane, and the relatively narrow redshift range in our analysis, the observed volume is small. Consequently, the most luminous (large) galaxies are sparse. The bright-end magnitude limit of our sample reaches only $M_{UV}=-19.5$ mag, with just 1 and 6 sources in the two brightest luminosity intervals of [-20,-19] and [-19,-18], respectively. 
Given the limited coverage of our sample at the bright end, we do not expect to effectively constrain $M^*$ simultaneously with the other two Schechter parameters. To address this, we apply a Gaussian prior of $M^* = -18.5\pm0.1(1 \sigma)$, adopted from the UVCANDELS results at a similar redshift \citep{Sun.2024}. In addition, a flat prior of $-3<\alpha<0$ is applied throughout the analysis. We quantify uncertainties in our best-fit parameters using MCMC analysis, implemented through the Python package \texttt{emcee} \citep{Foreman13}. Note that in the MLE fitting technique, the normalization parameter of the Schechter function cancels out and is not directly fitted. We therefore estimate it separately based on the number counts \citep[e.g.,][]{Sun.2024}.

\begin{figure*}
\centering
\includegraphics[width=0.9\textwidth]{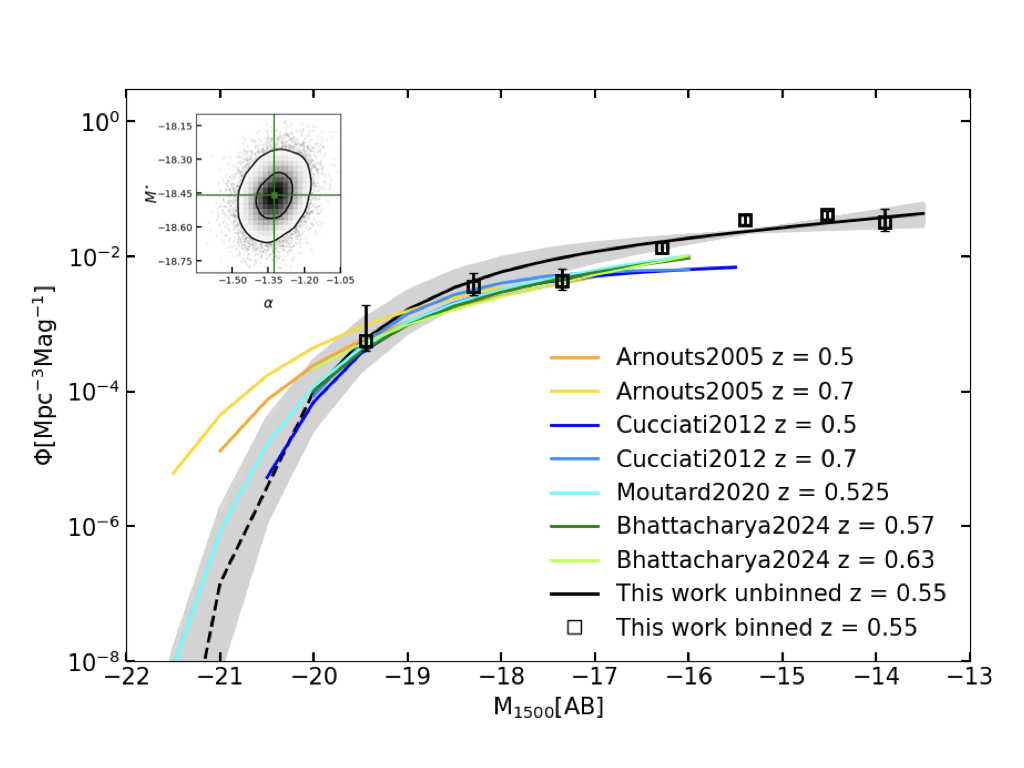}
\caption{Rest-frame UV LF for galaxies at $z\sim0.55$ in the Abell 2744 field. The best-fit Schechter function from our MLE is plotted as a black solid curve, with the surrounding gray band indicating its $3\sigma$ uncertainty range. For reference, a black dashed line extends our best-fit model to brighter magnitudes, where direct constraints from our data are absent, to aid comparison with other studies. The observed, binned LF (shown as black squares) is provided for visual assessment only. Various literature measurements at similar redshifts are overplotted as colored lines, as detailed in the legend. The inset panel presents the joint constraints on the Schechter parameters, the faint-end slope $\alpha$ and the characteristic magnitude $M^{*}$. Green lines mark the median (best-fit) values reported in Table~\ref{tab:pars}, while the black contours enclose the $1\sigma$ (0.393) and $2\sigma$ (0.865) confidence regions.
}
\label{fig:LF}
\end{figure*}

Figure~\ref{fig:LF} presents the rest-frame UV LF for our galaxy sample in the Abell 2744 field, derived at a median redshift of $z \sim 0.55$ using the MLE technique detailed previously. The best-fit Schechter function is shown as a black solid curve. A black dashed line extends the best-fit model to brighter magnitudes beyond our direct observational constraints, facilitating comparison with other studies that probe the bright end of the LF. To quantify the LF uncertainties, we evaluate the model luminosity function at each $M_{\mathrm{UV}}$ using an ensemble of $(\alpha, M^{*})$ parameter pairs drawn from bootstrapped MCMC chains. The resulting $3\sigma$ confidence region is represented by the surrounding gray shaded band. The resulting best-fit Schechter parameters are listed in Table~\ref{tab:pars}. The inset panel illustrates the joint posterior distribution of the faint-end slope $\alpha$ and the characteristic magnitude $M^{*}$, with green lines marking their median (best-fit) values and black contours outlining the $1\sigma$ and $2\sigma$ confidence regions.

We complement the unbinned maximum likelihood fit in Figure~\ref{fig:LF} with a binned representation of the luminosity function (black squares), which includes Poissonian uncertainties and serves primarily for visual comparison. The binned LF is computed following \citet{Alavi.2016} using the $1/V_{\mathrm{eff}}$ estimator, where the number density in each magnitude bin is obtained by summing the inverse effective volume $V_{\mathrm{eff}}(M_j)$ over all $N_i$ galaxies within that bin. Error bars for the binned data are derived from Poisson statistics, with appropriate corrections applied to bins containing fewer than 50 galaxies following the formulation of \citet{Gehrels86}. All bins have a fixed width of $\Delta M_{\mathrm{UV}} = 1$ mag. Because the redshift range of our sample is narrow, the variation of $V_{\mathrm{eff}}$ within a given magnitude bin is limited. This ensures that individual galaxies carry similar weights, thereby validating the use of simple Poissonian errors. We emphasize, however, that the binned LF is presented solely for visual assessment and should not be used for quantitative parameter estimation, as its value is sensitive to the chosen binning scheme and inevitably discards information present in the unbinned data.

Figure~\ref{fig:LF} compares our results with published rest-UV LFs at $\sim$1500\AA and similar redshifts, including data from GALEX \citep{Arnouts05}, VVDS \citep{Cucciati12}, CLAUDS and HSC-SSP \citep{Moutard20} and HST \citep{Bhattacharya.2024}. While our bright-end measurements agree with previous studies, we probe significantly fainter magnitudes. Our maximum likelihood analysis yields a faint-end slope of $\alpha = -1.324^{+0.072}_{-0.074}$ at $z\sim0.55$. This result agrees with \citet{Moutard20} and \citet{Bhattacharya.2024} at the $\sim1\sigma$ level, who measured $\alpha = -1.408^{+0.053}_{-0.053}$ at $z\sim0.525$ and $\alpha = -1.41^{+0.12}_{-0.12}$ at $z\sim0.57$, respectively. Our result is slightly flatter than other studies at similar redshifts, particularly those with brighter magnitude limits like \citet{Arnouts05}. 

Beyond the Poisson uncertainties reflected in the error bars of Figure~\ref{fig:LF}, the galaxy number counts in the Abell 2744 field are susceptible to cosmic variance due to the survey's limited volume. To assess this additional uncertainty, we employ the \texttt{Cosmic Variance Calculator v1.03}\footnote{\url{https://www.ph.unimelb.edu.au/~mtrenti/cvc/CosmicVariance.html}} \citep{Trenti08}, implementing the halo bias model of \citet{Sheth99} with parameter choices $\sigma_8 = 0.8$ and an average halo occupation fraction of $0.5$. For galaxies in our brightest magnitude bins, the fractional error introduced by cosmic variance is estimated to be approximately $0.2$. This systematic contribution is notably lower than the corresponding statistical Poisson uncertainty (around $0.7$ for the same bright population), indicating that Poisson noise remains the dominant source of error at the bright end of our luminosity function.

\begin{deluxetable*}{lllllll}
\tabletypesize{\footnotesize}
\tablecaption{Best-fit Schechter Parameters for the UV LFs and the UV Luminosity Density}  
\tablehead{ 
\colhead{Redshift} & \colhead{$M_{\mathrm{lim,UV}}(\mathrm{Mag})$} & \colhead{$N$\tablenotemark{a}} & \colhead{$\alpha$} & \colhead{$M^{*}(\mathrm{Mag})$} & \colhead{$\phi^{*} (10^{-3}\mathrm{Mpc}^{-3})$} & \colhead{$\rho_{\mathrm{UV}}$\tablenotemark{b}}
}
\startdata
$0.4<z<0.7$ & -13.5 & 147 & $-1.324^{+0.072}_{-0.074}$ & $-18.460^{+0.101}_{-0.099}$ & $10.66^{+2.54}_{-2.19}$  & $1.31^{+0.178}_{-0.203}$ \\
\enddata
\label{tab:pars}
\tablenotetext{a}{Sample size after removing sources with completeness $< 50\%$. }
\tablenotetext{b}{in units of $\times 10^{26}$ ergs/s/Hz/Mpc$^3$}
\end{deluxetable*}

\section{Discussion} \label{sec:discusstion}

\subsection{Evolution of the Faint-end Slope}\label{sec:evol}

\begin{figure}
\centering
\includegraphics[width=0.5\textwidth]{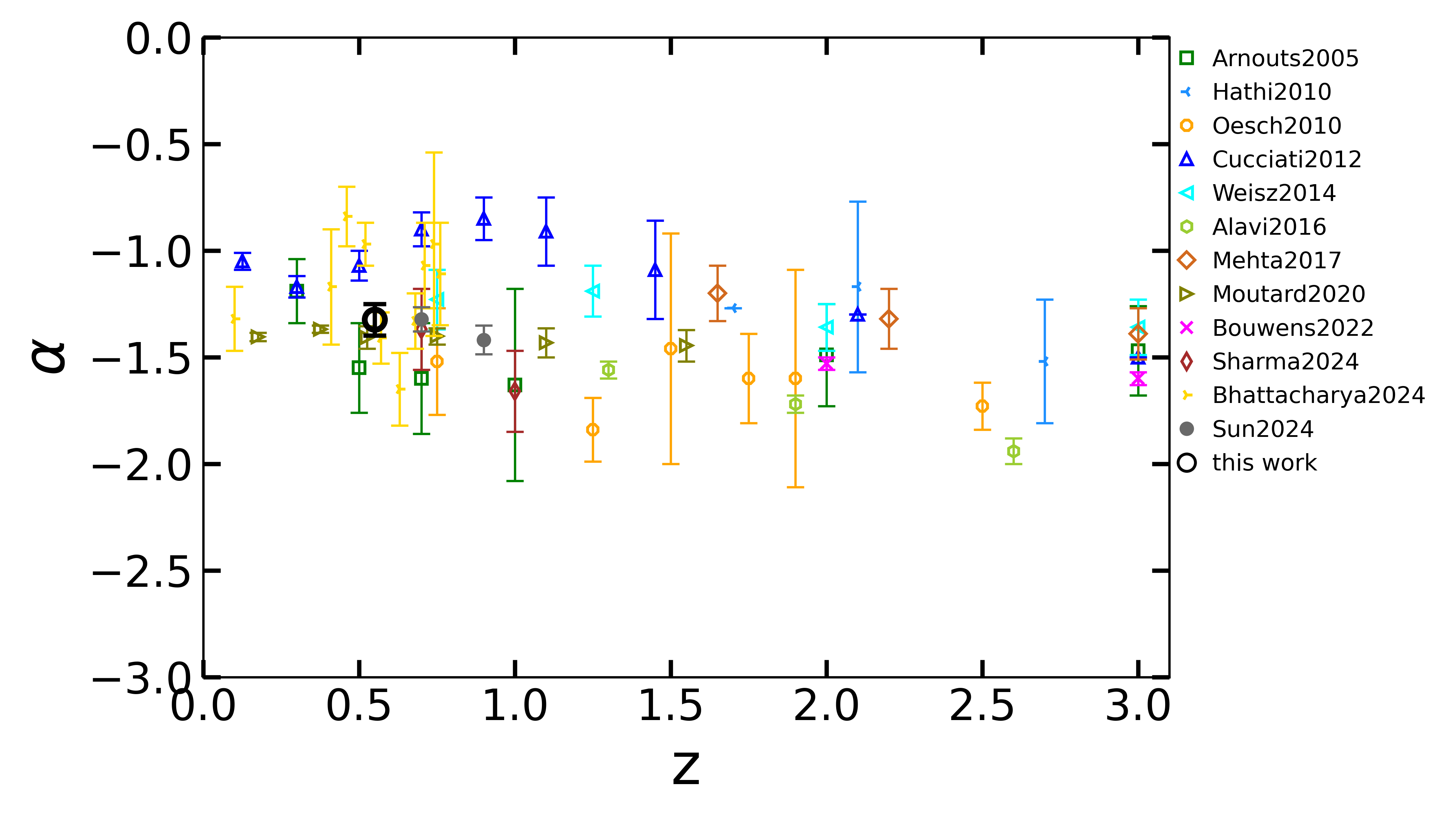}
\caption{Redshift evolution of the faint-end slope $\alpha$ of the LF. The black dot shows the result from A2744 F225W sample at $z\sim0.55$. Symbols representing previous determinations from the literature are summarized in the legend. The error bars show the $1\sigma$ uncertainties.}
\label{fig:par_evol}
\end{figure}

Since in our analysis $M^*$ is primarily constrained by the prior adopted from the UVCANDELS measurements at similar redshifts, we focus here on investigating the evolution of the fain-end slope $\alpha$ with redshift. We compile determinations of $\alpha$ across the redshift range $0<z<3$ from the literature \citep{Arnouts05,Hathi10,Cucciati12,Weisz14,Alavi16,Mehta.2017,Moutard20,Bouwens.2022c,Sharma.2024,Bhattacharya.2024,Sun.2024}, alongside our best-fit value in Figure~\ref{fig:par_evol}. The literature measurements included in the figure are predominantly at rest-frame $\sim1500$\AA, with the exception of the $1700$\AA\ results from \citet{Bouwens.2022c}. Among these results, both \cite{Alavi16} and \cite{Bouwens.2022c} utilized data in HFF cluster fields to investigate the faint end of UV LF, reaching $\sim-12$ mag. However, as their observations employed redder bands than the F225W, their measurements on $\alpha$ are primarily constrained to $z > 1$. At $z<1$, our results using the F225W observations in the HFF fields currently provide the only constraints directly reaching this faint luminosity limit.

Our best-fit faint-end slope ($\alpha= -1.324^{+0.072}_{-0.074}$ at $z\sim0.55$) is in line with other measurements obtained at similar redshifts, as seen in Figure~\ref{fig:par_evol} when comparing values across different epochs. Notably, our results show better agreement with the measurements of \citet{Moutard20} at $z=0.525$ and \citet{Bhattacharya.2024} at $z=0.57$. Combining our Abell 2744 measurement with the UVCANDELS blank-field results at $z\sim 0.7-0.9$ \citep{Sun.2024}, we identify an evolutionary trend: the faint-end slope $\alpha$ progressively steepens at higher redshifts. This trend is further strengthened when considered across the full $0<z<3$ redshift baseline shown in Figure~\ref{fig:par_evol}.

\subsection{UV Luminosity Density}\label{sec:uvld}

The total unobscured UV light emitted by galaxies, quantified by the cosmic UV luminosity density $\rho_\text{UV}$, is derived from our best-fit Schechter function (Section~\ref{sec:lf}). This quantity is calculated by integrating the product of luminosity and the LF over the full range of galaxy brightness:

\begin{equation}
\label{eq:uvld}
\rho_\text{UV} = \int_{L_\text{lim}}^{\infty} L\ \phi(L)\ dL= \int_{-\infty}^{M_\text{lim}} L(M)\ \phi(M)\ dM
\end{equation}

 In Table~\ref{tab:pars}, we present the cumulative UV luminosity density computed by integrating down to $M_\text{UV}=-15$. We adopt this magnitude limit based on findings from our subsequent analysis (Section~\ref{sec:turnover}), which suggests a potential turn-over in the UV LF fainter than $M_{\rm UV}\sim -15$. The uncertainty in $\rho_\mathrm{UV}$ is propagated from the fitted Schechter parameters ($\alpha$, $M^{*}$). We achieve this by first drawing a large sample of $(\alpha, M^{*})$ pairs from their two-dimensional posterior distribution via MCMC. For each sampled pair, we evaluate Equation~(\ref{eq:uvld}) to compute a corresponding $\rho_\mathrm{UV}$ value. The ensemble of these computed values forms a probability distribution for $\rho_\mathrm{UV}$, from which we directly infer its $1\sigma$ confidence interval.
 
\begin{figure}
\centering
\includegraphics[width=\columnwidth]{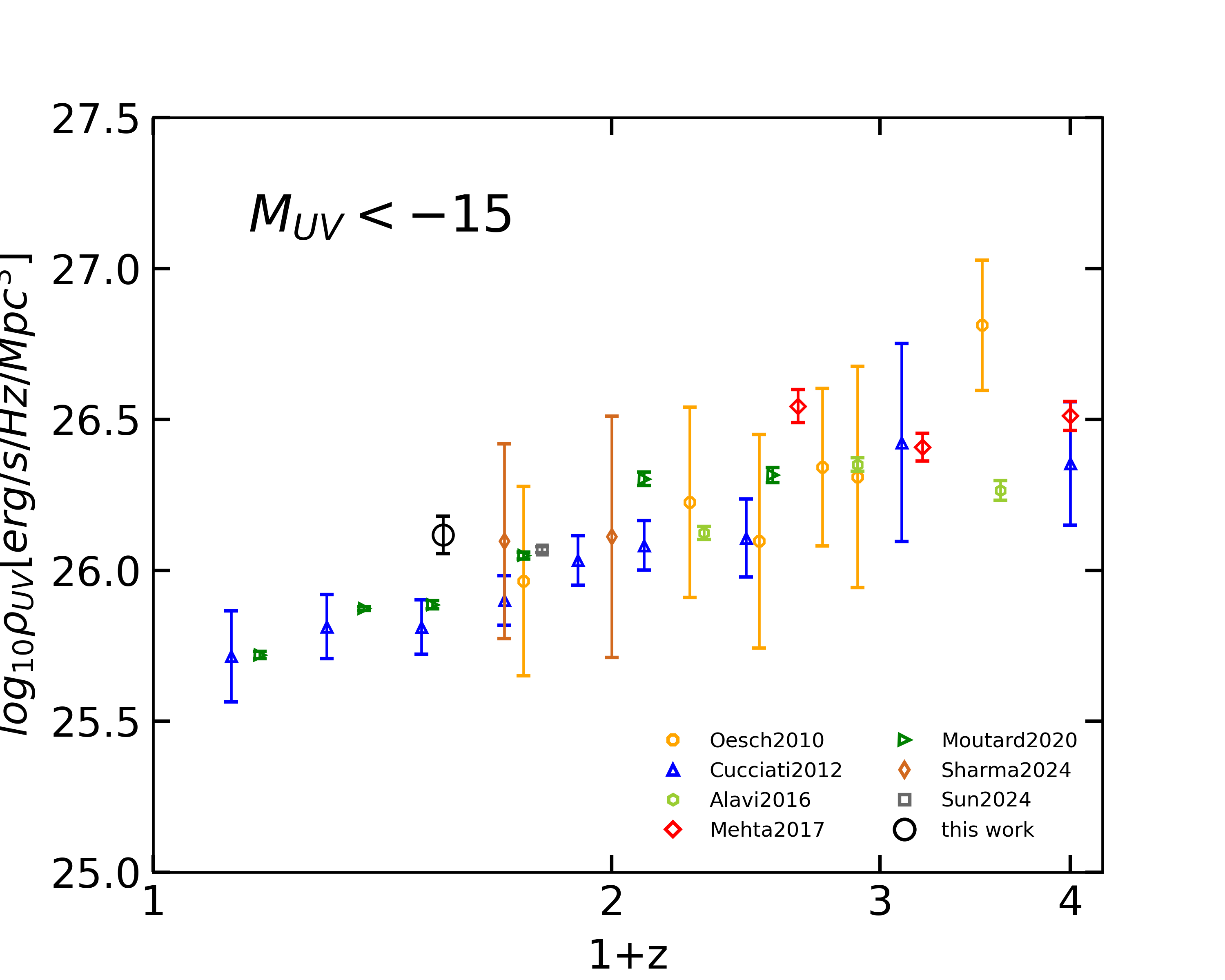}
\caption{Redshift evolution of the UV luminosity density. The black open circle represents the results computed from the best-fit Schechter parameters of our sample at $z\sim0.55$ (not corrected for dust). Symbols corresponding to results from the literature are summarized in the legend. All data points are obtained by integrating the rest-frame UV LFs down to $M_{UV}=-15$. The $1\sigma$ uncertainties are derived using the reported uncertainties of the LF parameters in each reference.}
\label{fig:rho_evol}
\end{figure}

 Figure~\ref{fig:rho_evol} presents the evolution of the UV luminosity density over the redshift range $0<z<3$. To ensure a uniform comparison, we have recomputed every literature $\rho_\text{UV}$ value by applying Equation~(\ref{eq:uvld}) with the authors’ published Schechter parameters, integrating in all cases to our adopted limit of $M_\text{UV} = -15$ \citep{Oesch10, Cucciati12, Alavi16, Mehta.2017, Moutard20, Sharma.2024,Sun.2024}. Note that uncertainty estimation of $\rho_\mathrm{UV}$ cannot be done straightforwardly, since covariance between Schechter parameters are usually not provided in the literature. Following \cite{Madau.2014}, we consequently employ fixed fractional uncertainties ($\Delta\rho_\mathrm{UV}/\rho_\mathrm{UV}$) reported in the original studies to derive consistent error estimates for $\rho_\mathrm{UV}$ values integrated to $M_\text{UV}=-15$. 

Figure~\ref{fig:rho_evol} shows that our derived $\rho_\text{UV}$ is consistent with \citet{Sharma.2024} within $1\sigma$ levels at comparable redshifts, though it is relatively higher than other literature values at adjacent redshifts. Taken together, a steady increase in the unobscured UV luminosity density is observed from $z=0$ to $z=2$, consistent with findings from several prior investigations (e.g., \citealt{Cucciati12}; \citealt{Alavi16};\citealt{Bhattacharya.2025}). Furthermore, work by \citet{Bouwens.2022c} indicates that even after accounting for dust attenuation, this rising trend persists over $0<z<2$. Considering the well-established proportionality between UV luminosity density and cosmic star formation rate (SFR) density, the observed evolution in $\rho_\text{UV}$ implies a steady increase in the global SFR from $z=0$ to $z\sim2$.

\subsection{Possible Turn-over in the LF}\label{sec:turnover}

The deep HFF observations in lensed fields have probed into very faint luminosities, which enables us to investigate a possible turn-over at the faint end of the UV LF. These investigations significantly advance our understanding of galaxy evolution by constraining star formation efficiency in low-mass galaxies and of cosmic reionization by enabling precise quantification of ionizing photon contributions from ultra-faint star-forming galaxies.

Following \citet{Bouwens.2017}, we extend the standard Schechter parameterization to allow for a possible departure from a pure power law at faint magnitudes. The modified form includes an additional curvature term in magnitude:
\begin{displaymath}
10^{-0.4\delta (M+16)^2}
\end{displaymath}
Here, the parameter $\delta$ controls the degree of curvature faintward of the pivot magnitude $M = -16$. Positive $\delta$ result in a turn-over, while negative $\delta$ cause the LF to steepen. The turnover magnitude $M_T$, where the derivative $d\phi/dM$ equals zero, is then expressed by the following relation:
\begin{equation}
M_T = -16 - \frac{\alpha+1}{2\delta}
\label{eq:mt}
\end{equation}
for $\delta > 0$. 

\begin{figure}
\centering
\includegraphics[width=0.45\textwidth]{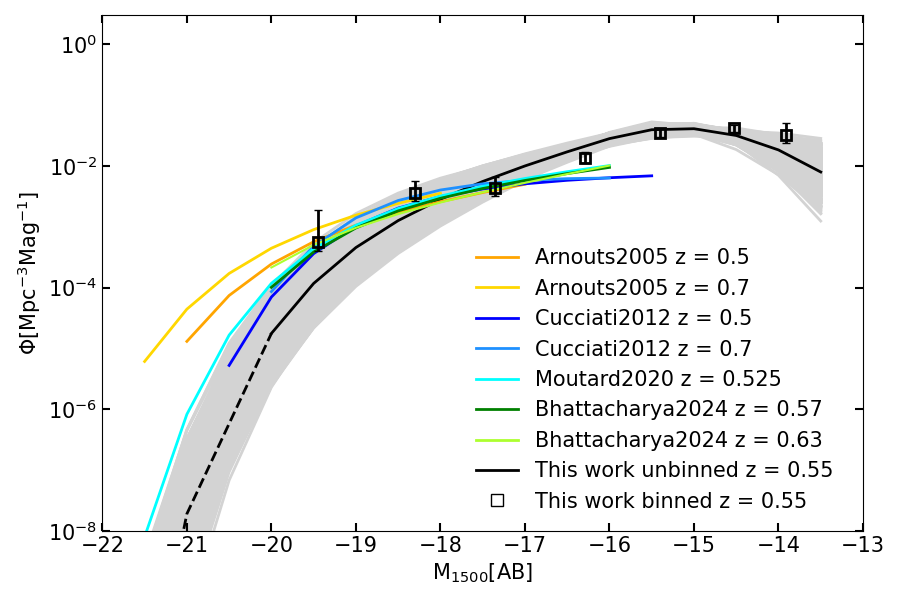}
\caption{Similar to Figure~\ref{fig:LF}, but incorporating a curvature parameter $\delta$ to account for a possible turn-over at the faint end of the UV LF.}
\label{fig:LF_delta}
\end{figure}

\begin{figure}
\centering
\includegraphics[width=0.45\textwidth]{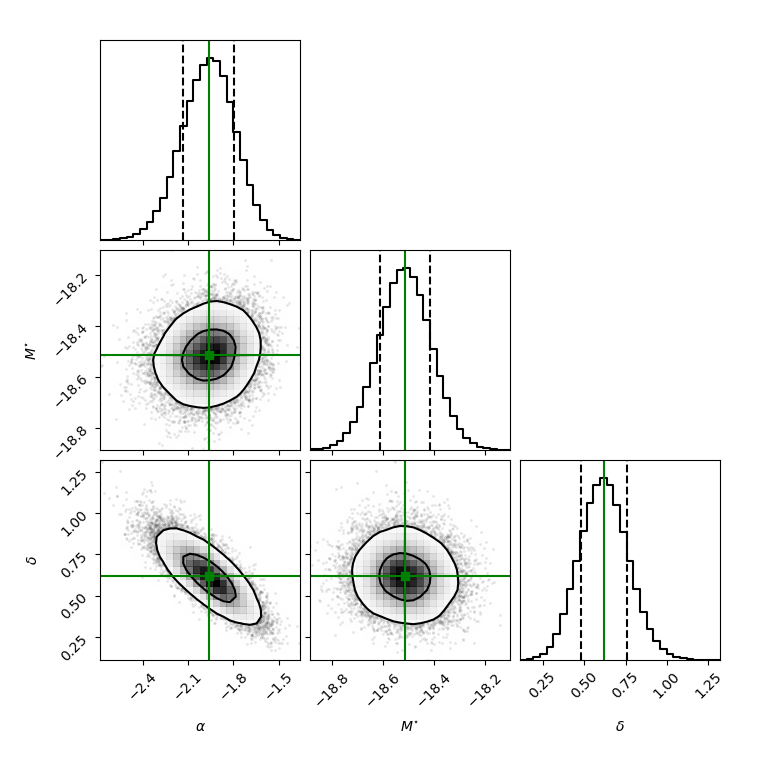}
\caption{Maximum likelihood estimates of the Schechter LF parameters $(\alpha,M^{*})$ and the curvature parameter $\delta$. The green lines label the best-fit (median) values. The black contours show the $1\sigma$(0.393) and $2\sigma$ ($0.865$) confidence levels.}
\label{fig:Corner}
\end{figure}

\begin{deluxetable}{lllll}
\tablecaption{Best-fit Schechter Parameters and Curvature Parameter for the UV LFs}  
\label{tab:par_delta}
\tablehead{ 
\colhead{$\alpha$} & \colhead{$M^{*}(\mathrm{Mag})$} & \colhead{$\phi^{*} (10^{-3}\mathrm{Mpc}^{-3})$} &  \colhead{$\delta$} & \colhead{$M_T$}
}
\startdata
$-1.961^{+0.164}_{-0.175}$ & $-18.514^{+0.099}_{-0.098}$ & $3.62^{+1.67}_{-1.24}$  & $0.619^{+0.141}_{-0.139}$ & $-15.218^{+0.120}_{-0.099}$ \\
\enddata
\end{deluxetable}

We incorporate this new parameterization of the LF into our maximum likelihood estimation. The best-fit UV LF and confidence intervals of the corresponding parameters are presented in Figure~\ref{fig:LF_delta} and~\ref{fig:Corner}, respectively, after incorporating the curvature parameter $\delta$ to account for a possible turn-over at the faint end. As also listed in Table~\ref{tab:par_delta}, we find a positive curvature of $\delta=0.619^{+0.141}_{-0.139}$, implying a possible turn-over at the magnitude of $M_T=-15.218^{+0.120}_{-0.099}$. Conservatively, our results rule out the existence of a turn-over in the UV LF brighter than -15.5 mag at the $3\sigma$ level. Although few previous studies have reached such faint magnitudes within the redshift range $z\sim0.5-1$ to investigate possible turn-overs of the UV LFs, our findings are in excellent agreement with studies at higher redshifts by \cite{Bouwens.2022c}, which rule out a turn-over in the UV LF brighter than -15.5 mag ($95\%$ confidence) over the redshift range $z = 2-9$. Additionally, previous studies by \cite{Atek.2015, Atek.2018}, \cite{Castellano.2016n}, \cite{Livermore.2017}, \cite{Bouwens.2017}, \cite{Yue.2018n}, \cite{Ishigaki18} and \cite{Bhatawdekar.2019} also report consistent results at $z > 2$. 

It is noted that when incorporating the curvature parameter $\delta$, the best-fit value of the faint-end slope changes significantly, from $-1.324^{+0.072}_{-0.074}$ to $-1.961^{+0.164}_{-0.175}$, as shown in Table~\ref{tab:par_delta}. From Figure~\ref{fig:LF_delta}, since the decline of the LF faintward -16 mag is accounted for by the $\delta$ parameter, $\alpha$ is more influenced by the rising trend brightward -16 mag. Intuitively, this results in a better fit to our data. We also estimate the Bayesian Information Criterion (BIC) for both cases and find that the BIC for the case with $\delta$ is $2\%$ smaller than that for the case without $\delta$. This further demonstrates that incorporating the curvature parameter $\delta$ to account for a possible turn-over at the faint end of the UV LF does provide a better fit to our data.

\section{Summary} \label{sec:sum}

We have obtained deep Near-UV F225W imaging of all six Hubble Frontier Fields clusters through the HST Cycle-27 program GO15940. In this first paper of a series, we present the data reduction and photometric analysis of the Abell 2744 F225W observations. By leveraging advantages of both deep HST imaging and the strong gravitational lensing magnification, we identify a sample of 152 ultra-faint galaxies with $M_{UV}<-12.1$ at $z\sim0.55$ in the Abell 2744 field, selected through a hybrid spectroscopic/photometric redshift approach. We conduct a series of simulations using \texttt{GLACiAR2+}, an image-based injection-recovery package specifically adapted for lensing fields, to calculate the completeness corrections necessary for robust LF measurements. We determine the UV LF at $z\sim0.55$ to its faint end and examine its evolutionary trends across redshifts through comparison with literature results.

We derive the best-fit Schechter UV LF using a maximum likelihood technique considering various sources 
of uncertainty including the lensing models. Gravitational lensing magnification helps our measurements to reach a faint magnitude of $M_{UV}=-13.5$ (after a $50\%$ completeness cut) at $0.4<z<0.7$, yielding a robust faint-end slope measurement of  $\alpha =-1.324^{+0.072}_{-0.074}$. This measurement is in consistency with previous studies of \citet{Moutard20} and \citet{Bhattacharya.2024} at the $\sim 1\sigma$ level. Our analysis, when integrated with previous UVCANDELS blank-field results, indicates a gradual evolution toward steeper faint-end slopes at higher redshifts. This evolutionary trend becomes increasingly pronounced when analyzed across more extended redshift baselines.

We obtain an unobscured cumulative UV luminosity density down to $M_\text{UV}<-15$ assuming there is no turn-over in the UV LFs till this limit. Our estimated $\rho_\text{UV}$ using the Abell 2744 F225W sample at $z\sim0.55$ is in good agreement with results from \citet{Sharma.2024} at $z=0.7$. The observed steepening of the faint-end slope of the UV LF with redshift suggests an increasing contribution from low-luminosity galaxies to the total cosmic UV luminosity density at higher redshifts. When combined with studies of Lyman-continuum escape fractions, it can help determine whether faint galaxies serve as the dominant source of ionizing photons during cosmic reionization.

This study highlights the unique power of gravitational lensing to provide robust constraints on the faint end of the luminosity function. Our new measurements enable constraints on a potential turn-over in the UV LF, and we rule out the existence of a turn-over brighter than -15.5 mag at the $3\sigma$ confidence level at $z\sim 0.55$. This is a progress in the faint-end turn-over studies, as few previous investigations have directly probed such low luminosities ($M_{\rm UV}\sim-13.5$) at redshifts below $z=1$. 

\begin{acknowledgments}
We thank the referee for the insightful and constructive comments, which have significantly improved the quality of this manuscript. We thank Michele Trenti, Yuxuan Pang and Mengting Ju for helpful discussion.
This work is supported by the China Manned Space Program with grant no. CMS-CSST-2025-A06, the National Key R\&D Program of China No.2025YFF0510603, the National Natural Science Foundation of China (grant 12373009), the CAS Project for Young Scientists in Basic Research Grant No. YSBR-062, and the Fundamental Research Funds for the Central Universities. XW acknowledges the support by the Xiaomi Young Talents Program, and the work carried out, in part, at the Swinburne University of Technology, sponsored by the ACAMAR visiting fellowship.
This work is based on observations with the NASA/ESA Hubble Space Telescope obtained at the Space Telescope Science Institute, which is operated by the Association of Universities for Research in Astronomy, Incorporated, under NASA contract NAS5-26555. 
This work utilizes gravitational lensing models produced by PIs Bradač, Natarajan \& Kneib (CATS), Merten \& Zitrin, Sharon, Williams, Keeton, Bernstein and Diego, and the GLAFIC group. This lens modeling was partially funded by the HST Frontier Fields program conducted by STScI. STScI is operated by the Association of Universities for Research in Astronomy, Inc. under NASA contract NAS 5-26555. The lens models were obtained from the Mikulski Archive for Space Telescopes (MAST).
\end{acknowledgments}


\begin{thebibliography}{}
\expandafter\ifx\csname natexlab\endcsname\relax\def\natexlab#1{#1}\fi
\providecommand{\url}[1]{\href{#1}{#1}}
\providecommand{\dodoi}[1]{doi:~\href{http://doi.org/#1}{\nolinkurl{#1}}}
\providecommand{\doeprint}[1]{\href{http://ascl.net/#1}{\nolinkurl{http://ascl.net/#1}}}
\providecommand{\doarXiv}[1]{\href{https://arxiv.org/abs/#1}{\nolinkurl{https://arxiv.org/abs/#1}}}

\bibitem[{{Adak} {et~al.}(2024){Adak}, {Hazra}, {Mitra}, \& {Krishak}}]{Adak.2024}
{Adak}, D., {Hazra}, D.~K., {Mitra}, S., \& {Krishak}, A. 2024, \jcap, 2024, 010, \dodoi{10.1088/1475-7516/2024/12/010}

\bibitem[{{Adams} {et~al.}(2020){Adams}, {Bowler}, {Jarvis}, {H{\"a}u{\ss}ler}, {McLure}, {Bunker}, {Dunlop}, \& {Verma}}]{Adams20}
{Adams}, N.~J., {Bowler}, R.~A.~A., {Jarvis}, M.~J., {et~al.} 2020, \mnras, 494, 1771, \dodoi{10.1093/mnras/staa687}

\bibitem[{{Adams} {et~al.}(2023){Adams}, {Bowler}, {Jarvis}, {Varadaraj}, \& {H{\"a}u{\ss}ler}}]{Adams23a}
{Adams}, N.~J., {Bowler}, R.~A.~A., {Jarvis}, M.~J., {Varadaraj}, R.~G., \& {H{\"a}u{\ss}ler}, B. 2023, \mnras, 523, 327, \dodoi{10.1093/mnras/stad1333}

\bibitem[{{Adams} {et~al.}(2024){Adams}, {Conselice}, {Austin}, {Harvey}, {Ferreira}, {Trussler}, {Juod{\v{z}}balis}, {Li}, {Windhorst}, {Cohen}, {Jansen}, {Summers}, {Tompkins}, {Driver}, {Robotham}, {D'Silva}, {Yan}, {Coe}, {Frye}, {Grogin}, {Koekemoer}, {Marshall}, {Pirzkal}, {Ryan}, {Maksym}, {Rutkowski}, {Willmer}, {Hammel}, {Nonino}, {Bhatawdekar}, {Wilkins}, {Bradley}, {Broadhurst}, {Cheng}, {Dole}, {Hathi}, \& {Zitrin}}]{Adams.2024}
{Adams}, N.~J., {Conselice}, C.~J., {Austin}, D., {et~al.} 2024, \apj, 965, 169, \dodoi{10.3847/1538-4357/ad2a7b}

\bibitem[{Alavi {et~al.}(2014)Alavi, Siana, Richard, Stark, Scarlata, Teplitz, Freeman, Domínguez, Rafelski, Robertson, \& Kewley}]{Alavi.2014}
Alavi, A., Siana, B.~D., Richard, J., {et~al.} 2014, ApJ, 780, 143, \dodoi{10.1088/0004-637x/780/2/143}

\bibitem[{{Alavi} {et~al.}(2016){Alavi}, {Siana}, {Richard}, {Rafelski}, {Jauzac}, {Limousin}, {Freeman}, {Scarlata}, {Robertson}, {Stark}, {Teplitz}, \& {Desai}}]{Alavi16}
{Alavi}, A., {Siana}, B., {Richard}, J., {et~al.} 2016, \apj, 832, 56, \dodoi{10.3847/0004-637X/832/1/56}

\bibitem[{Alavi {et~al.}(2016)Alavi, Siana, Richard, Rafelski, Jauzac, Limousin, Freeman, Scarlata, Robertson, Stark, Teplitz, \& Desai}]{Alavi.2016}
Alavi, A., Siana, B., Richard, J., {et~al.} 2016, The Astrophysical Journal, 832, 56, \dodoi{10.3847/0004-637x/832/1/56}

\bibitem[{{Ammons} {et~al.}(2014){Ammons}, {Wong}, {Zabludoff}, \& {Keeton}}]{Ammons.2014}
{Ammons}, S.~M., {Wong}, K.~C., {Zabludoff}, A.~I., \& {Keeton}, C.~R. 2014, \apj, 781, 2, \dodoi{10.1088/0004-637X/781/1/2}

\bibitem[{{Anderson} {et~al.}(2021){Anderson}, {Baggett}, \& {Kuhn}}]{Anderson.2021}
{Anderson}, J., {Baggett}, S., \& {Kuhn}, B. 2021, Instrument Science Report 2021-9, 44 pages

\bibitem[{{Arnouts} {et~al.}(2005){Arnouts}, {Schiminovich}, {Ilbert}, {Tresse}, {Milliard}, {Treyer}, {Bardelli}, {Budavari}, {Wyder}, {Zucca}, {Le F{\`e}vre}, {Martin}, {Vettolani}, {Adami}, {Arnaboldi}, {Barlow}, {Bianchi}, {Bolzonella}, {Bottini}, {Byun}, {Cappi}, {Charlot}, {Contini}, {Donas}, {Forster}, {Foucaud}, {Franzetti}, {Friedman}, {Garilli}, {Gavignaud}, {Guzzo}, {Heckman}, {Hoopes}, {Iovino}, {Jelinsky}, {Le Brun}, {Lee}, {Maccagni}, {Madore}, {Malina}, {Marano}, {Marinoni}, {McCracken}, {Mazure}, {Meneux}, {Merighi}, {Morrissey}, {Neff}, {Paltani}, {Pell{\`o}}, {Picat}, {Pollo}, {Pozzetti}, {Radovich}, {Rich}, {Scaramella}, {Scodeggio}, {Seibert}, {Siegmund}, {Small}, {Szalay}, {Welsh}, {Xu}, {Zamorani}, \& {Zanichelli}}]{Arnouts05}
{Arnouts}, S., {Schiminovich}, D., {Ilbert}, O., {et~al.} 2005, \apjl, 619, L43, \dodoi{10.1086/426733}

\bibitem[{Atek {et~al.}(2018)Atek, Richard, Kneib, \& Schaerer}]{Atek.2018}
Atek, H., Richard, J., Kneib, J.-P., \& Schaerer, D. 2018, MNRAS, 479, 5184 , \dodoi{10.1093/mnras/sty1820}

\bibitem[{Atek {et~al.}(2014)Atek, Kneib, Pacifici, Malkan, Charlot, Lee, Bedregal, Bunker, Colbert, Dressler, Hathi, Lehnert, Martin, McCarthy, Rafelski, Ross, Siana, \& Teplitz}]{Atek.2014}
Atek, H., Kneib, J.-P., Pacifici, C., {et~al.} 2014, ApJ, 789, 96, \dodoi{10.1088/0004-637x/789/2/96}

\bibitem[{Atek {et~al.}(2015)Atek, Richard, Kneib, Jauzac, Schaerer, Clement, Limousin, Jullo, Natarajan, Egami, \& Ebeling}]{Atek.2015}
Atek, H., Richard, J., Kneib, J.-P., {et~al.} 2015, ApJ, 800, 18, \dodoi{10.1088/0004-637x/800/1/18}

\bibitem[{{Bagley} {et~al.}(2024){Bagley}, {Finkelstein}, {Rojas-Ruiz}, {Diekmann}, {Finkelstein}, {Song}, {Papovich}, {Somerville}, {Baronchelli}, \& {Dai}}]{Bagley.2024}
{Bagley}, M.~B., {Finkelstein}, S.~L., {Rojas-Ruiz}, S., {et~al.} 2024, \apj, 961, 209, \dodoi{10.3847/1538-4357/ad09dc}

\bibitem[{{Bartelmann}(2010)}]{Bartelmann.2010}
{Bartelmann}, M. 2010, Classical and Quantum Gravity, 27, 233001, \dodoi{10.1088/0264-9381/27/23/233001}

\bibitem[{{Belles} {et~al.}(2025){Belles}, {Gronwall}, {Siegel}, {Ciardullo}, \& {Page}}]{Belles.2025}
{Belles}, A., {Gronwall}, C., {Siegel}, M.~H., {Ciardullo}, R., \& {Page}, M.~J. 2025, \apj, 979, 173, \dodoi{10.3847/1538-4357/ad9b2f}

\bibitem[{{Benson} {et~al.}(2003){Benson}, {Frenk}, {Baugh}, {Cole}, \& {Lacey}}]{Benson03}
{Benson}, A.~J., {Frenk}, C.~S., {Baugh}, C.~M., {Cole}, S., \& {Lacey}, C.~G. 2003, \mnras, 343, 679, \dodoi{10.1046/j.1365-8711.2003.06709.x}

\bibitem[{Bertin \& Arnouts(1996)}]{Bertin.1996}
Bertin, E., \& Arnouts, S. 1996, Astronomy and Astrophysics Supplement Series, 117, 393 , \dodoi{10.1051/aas:1996164}

\bibitem[{Bhatawdekar {et~al.}(2019)Bhatawdekar, Conselice, Margalef-Bentabol, \& Duncan}]{Bhatawdekar.2019}
Bhatawdekar, R., Conselice, C.~J., Margalef-Bentabol, B., \& Duncan, K. 2019, MNRAS, 486, 3805 , \dodoi{10.1093/mnras/stz866}

\bibitem[{{Bhattacharya} \& {Saha}(2025)}]{Bhattacharya.2025}
{Bhattacharya}, S., \& {Saha}, K. 2025, \mnras, 540, L65, \dodoi{10.1093/mnrasl/slaf036}

\bibitem[{Bhattacharya {et~al.}(2024)Bhattacharya, Saha, \& Mondal}]{Bhattacharya.2024}
Bhattacharya, S., Saha, K., \& Mondal, C. 2024, Monthly Notices of the Royal Astronomical Society, 532, 1059, \dodoi{10.1093/mnras/stae1583}

\bibitem[{Bouwens {et~al.}(2022)Bouwens, Illingworth, Ellis, Oesch, Paulino-Afonso, Ribeiro, \& Stefanon}]{Bouwens.2022}
Bouwens, R.~J., Illingworth, G., Ellis, R.~S., {et~al.} 2022, The Astrophysical Journal, 931, 81, \dodoi{10.3847/1538-4357/ac618c}

\bibitem[{{Bouwens} {et~al.}(2022{\natexlab{a}}){Bouwens}, {Illingworth}, {Ellis}, {Oesch}, \& {Stefanon}}]{Bouwens.2022c}
{Bouwens}, R.~J., {Illingworth}, G., {Ellis}, R.~S., {Oesch}, P., \& {Stefanon}, M. 2022{\natexlab{a}}, \apj, 940, 55, \dodoi{10.3847/1538-4357/ac86d1}

\bibitem[{{Bouwens} {et~al.}(2022{\natexlab{b}}){Bouwens}, {Illingworth}, {van Dokkum}, {Oesch}, {Stefanon}, \& {Ribeiro}}]{Bouwens.2022a}
{Bouwens}, R.~J., {Illingworth}, G.~D., {van Dokkum}, P.~G., {et~al.} 2022{\natexlab{b}}, \apj, 927, 81, \dodoi{10.3847/1538-4357/ac4791}

\bibitem[{Bouwens {et~al.}(2017)Bouwens, Oesch, Illingworth, Ellis, \& Stefanon}]{Bouwens.2017}
Bouwens, R.~J., Oesch, P.~A., Illingworth, G.~D., Ellis, R.~S., \& Stefanon, M. 2017, ApJ, 843, 129, \dodoi{10.3847/1538-4357/aa70a4}

\bibitem[{{Bouwens} {et~al.}(2021){Bouwens}, {Oesch}, {Stefanon}, {Illingworth}, {Labb{\'e}}, {Reddy}, {Atek}, {Montes}, {Naidu}, {Nanayakkara}, {Nelson}, \& {Wilkins}}]{Bouwens21}
{Bouwens}, R.~J., {Oesch}, P.~A., {Stefanon}, M., {et~al.} 2021, \aj, 162, 47, \dodoi{10.3847/1538-3881/abf83e}

\bibitem[{{Bowler} {et~al.}(2020){Bowler}, {Jarvis}, {Dunlop}, {McLure}, {McLeod}, {Adams}, {Milvang-Jensen}, \& {McCracken}}]{Bowler20}
{Bowler}, R.~A.~A., {Jarvis}, M.~J., {Dunlop}, J.~S., {et~al.} 2020, \mnras, 493, 2059, \dodoi{10.1093/mnras/staa313}

\bibitem[{Bradac {et~al.}(2005)Bradac, Schneider, Lombardi, \& Erren}]{Bradac.2005}
Bradac, M., Schneider, P., Lombardi, M., \& Erren, T.~C. 2005, A\&A, 437, 39 , \dodoi{10.1051/0004-6361:20042233}

\bibitem[{Bradac {et~al.}(2009)Bradac, Treu, Applegate, Gonzalez, Clowe, Forman, Jones, Marshall, Schneider, \& Zaritsky}]{Bradac.2009}
Bradac, M., Treu, T.~L., Applegate, D., {et~al.} 2009, ApJ, 706, 1201 , \dodoi{10.1088/0004-637x/706/2/1201}

\bibitem[{{Calamida} {et~al.}(2022){Calamida}, {Bajaj}, {Mack}, {Marinelli}, {Medina}, {Pidgeon}, {Kozhurina-Platais}, {Shanahan}, \& {Som}}]{2022AJ....164...32C}
{Calamida}, A., {Bajaj}, V., {Mack}, J., {et~al.} 2022, \aj, 164, 32, \dodoi{10.3847/1538-3881/ac73f0}

\bibitem[{Caminha {et~al.}(2017)Caminha, Grillo, Rosati, Meneghetti, Mercurio, Ettori, Balestra, Biviano, Umetsu, Vanzella, Annunziatella, Bonamigo, Delgado-Correal, Girardi, Lombardi, Nonino, Sartoris, Tozzi, Bartelmann, Bradley, Caputi, Coe, Ford, Fritz, Gobat, Postman, Seitz, \& Zitrin}]{Caminha.2017}
Caminha, G.~B., Grillo, C., Rosati, P., {et~al.} 2017, A\&A, 607, A93, \dodoi{10.1051/0004-6361/201731498}

\bibitem[{{Carrasco} {et~al.}(2018){Carrasco}, {Trenti}, {Mutch}, \& {Oesch}}]{Carrasco.2018}
{Carrasco}, D., {Trenti}, M., {Mutch}, S., \& {Oesch}, P.~A. 2018, \pasa, 35, e022, \dodoi{10.1017/pasa.2018.17}

\bibitem[{Casertano {et~al.}(2000)Casertano, Mello, Dickinson, Ferguson, Fruchter, Gonzalez-Lopezlira, Heyer, Hook, Levay, Lucas, Mack, Makidon, Mutchler, Smith, Stiavelli, Wiggs, \& Williams}]{Casertano.2000}
Casertano, S., Mello, D. F.~d., Dickinson, M.~E., {et~al.} 2000, AJ, 120, 2747 , \dodoi{10.1086/316851}

\bibitem[{{Castellano} {et~al.}(2016){Castellano}, {Yue}, {Ferrara}, {Merlin}, {Fontana}, {Amor{\'\i}n}, {Grazian}, {M{\'a}rmol-Queralto}, {Micha{\l}owski}, {Mortlock}, {Paris}, {Parsa}, {Pilo}, \& {Santini}}]{Castellano.2016n}
{Castellano}, M., {Yue}, B., {Ferrara}, A., {et~al.} 2016, \apjl, 823, L40, \dodoi{10.3847/2041-8205/823/2/L40}

\bibitem[{{Cucciati} {et~al.}(2012){Cucciati}, {Tresse}, {Ilbert}, {Le F{\`e}vre}, {Garilli}, {Le Brun}, {Cassata}, {Franzetti}, {Maccagni}, {Scodeggio}, {Zucca}, {Zamorani}, {Bardelli}, {Bolzonella}, {Bielby}, {McCracken}, {Zanichelli}, \& {Vergani}}]{Cucciati12}
{Cucciati}, O., {Tresse}, L., {Ilbert}, O., {et~al.} 2012, \aap, 539, A31, \dodoi{10.1051/0004-6361/201118010}

\bibitem[{Diego {et~al.}(2015)Diego, Broadhurst, Benítez, Lim, \& Lam}]{Diego.2015}
Diego, J.~M., Broadhurst, T., Benítez, N., Lim, J., \& Lam, D. 2015, MNRAS, 449, 588 , \dodoi{10.1093/mnras/stv298}

\bibitem[{{Diego} {et~al.}(2005{\natexlab{a}}){Diego}, {Protopapas}, {Sandvik}, \& {Tegmark}}]{Diego.2005a}
{Diego}, J.~M., {Protopapas}, P., {Sandvik}, H.~B., \& {Tegmark}, M. 2005{\natexlab{a}}, \mnras, 360, 477, \dodoi{10.1111/j.1365-2966.2005.09021.x}

\bibitem[{{Diego} {et~al.}(2005{\natexlab{b}}){Diego}, {Sandvik}, {Protopapas}, {Tegmark}, {Ben{\'\i}tez}, \& {Broadhurst}}]{Diego.2005b}
{Diego}, J.~M., {Sandvik}, H.~B., {Protopapas}, P., {et~al.} 2005{\natexlab{b}}, \mnras, 362, 1247, \dodoi{10.1111/j.1365-2966.2005.09372.x}

\bibitem[{{Diego} {et~al.}(2007){Diego}, {Tegmark}, {Protopapas}, \& {Sandvik}}]{Diego.2007}
{Diego}, J.~M., {Tegmark}, M., {Protopapas}, P., \& {Sandvik}, H.~B. 2007, \mnras, 375, 958, \dodoi{10.1111/j.1365-2966.2007.11380.x}

\bibitem[{{Donnan} {et~al.}(2023){Donnan}, {McLeod}, {Dunlop}, {McLure}, {Carnall}, {Begley}, {Cullen}, {Hamadouche}, {Bowler}, {Magee}, {McCracken}, {Milvang-Jensen}, {Moneti}, \& {Targett}}]{Donnan23}
{Donnan}, C.~T., {McLeod}, D.~J., {Dunlop}, J.~S., {et~al.} 2023, \mnras, 518, 6011, \dodoi{10.1093/mnras/stac3472}

\bibitem[{{Donnan} {et~al.}(2024){Donnan}, {McLure}, {Dunlop}, {McLeod}, {Magee}, {Arellano-C{\'o}rdova}, {Barrufet}, {Begley}, {Bowler}, {Carnall}, {Cullen}, {Ellis}, {Fontana}, {Illingworth}, {Grogin}, {Hamadouche}, {Koekemoer}, {Liu}, {Mason}, {Santini}, \& {Stanton}}]{Donnan.2024}
{Donnan}, C.~T., {McLure}, R.~J., {Dunlop}, J.~S., {et~al.} 2024, \mnras, 533, 3222, \dodoi{10.1093/mnras/stae2037}

\bibitem[{{Finkelstein} {et~al.}(2022{\natexlab{a}}){Finkelstein}, {Bagley}, {Song}, {Larson}, {Papovich}, {Dickinson}, {Finkelstein}, {Koekemoer}, {Pirzkal}, {Somerville}, {Yung}, {Behroozi}, {Ferguson}, {Giavalisco}, {Grogin}, {Hathi}, {Hutchison}, {Jung}, {Kocevski}, {Kawinwanichakij}, {Rojas-Ruiz}, {Ryan}, {Snyder}, \& {Tacchella}}]{Finkelstein22a}
{Finkelstein}, S.~L., {Bagley}, M., {Song}, M., {et~al.} 2022{\natexlab{a}}, \apj, 928, 52, \dodoi{10.3847/1538-4357/ac3aed}

\bibitem[{{Finkelstein} {et~al.}(2022{\natexlab{b}}){Finkelstein}, {Bagley}, {Arrabal Haro}, {Dickinson}, {Ferguson}, {Kartaltepe}, {Papovich}, {Burgarella}, {Kocevski}, {Huertas-Company}, {Iyer}, {Koekemoer}, {Larson}, {P{\'e}rez-Gonz{\'a}lez}, {Rose}, {Tacchella}, {Wilkins}, {Chworowsky}, {Medrano}, {Morales}, {Somerville}, {Yung}, {Fontana}, {Giavalisco}, {Grazian}, {Grogin}, {Kewley}, {Kirkpatrick}, {Kurczynski}, {Lotz}, {Pentericci}, {Pirzkal}, {Ravindranath}, {Ryan}, {Trump}, {Yang}, {Almaini}, {Amor{\'\i}n}, {Annunziatella}, {Backhaus}, {Barro}, {Behroozi}, {Bell}, {Bhatawdekar}, {Bisigello}, {Bromm}, {Buat}, {Buitrago}, {Calabr{\`o}}, {Casey}, {Castellano}, {Ch{\'a}vez Ortiz}, {Ciesla}, {Cleri}, {Cohen}, {Cole}, {Cooke}, {Cooper}, {Cooray}, {Costantin}, {Cox}, {Croton}, {Daddi}, {Dav{\'e}}, {de La Vega}, {Dekel}, {Elbaz}, {Estrada-Carpenter}, {Faber}, {Fern{\'a}ndez}, {Finkelstein}, {Freundlich}, {Fujimoto}, {Garc{\'\i}a-Argum{\'a}nez}, {Gardner}, {Gawiser}, {G{\'o}mez-Guijarro}, {Guo}, {Hamblin},
  {Hamilton}, {Hathi}, {Holwerda}, {Hirschmann}, {Hutchison}, {Jaskot}, {Jha}, {Jogee}, {Juneau}, {Jung}, {Kassin}, {Bail}, {Leung}, {Lucas}, {Magnelli}, {Mantha}, {Matharu}, {McGrath}, {McIntosh}, {Merlin}, {Mobasher}, {Newman}, {Nicholls}, {Pandya}, {Rafelski}, {Ronayne}, {Santini}, {Seill{\'e}}, {Shah}, {Shen}, {Simons}, {Snyder}, {Stanway}, {Straughn}, {Teplitz}, {Vanderhoof}, {Vega-Ferrero}, {Wang}, {Weiner}, {Willmer}, {Wuyts}, {Zavala}, \& {Ceers Team}}]{Finkelstein22b}
{Finkelstein}, S.~L., {Bagley}, M.~B., {Arrabal Haro}, P., {et~al.} 2022{\natexlab{b}}, \apjl, 940, L55, \dodoi{10.3847/2041-8213/ac966e}

\bibitem[{{Finkelstein} {et~al.}(2023){Finkelstein}, {Bagley}, {Ferguson}, {Wilkins}, {Kartaltepe}, {Papovich}, {Yung}, {Arrabal Haro}, {Behroozi}, {Dickinson}, {Kocevski}, {Koekemoer}, {Larson}, {Le Bail}, {Morales}, {P{\'e}rez-Gonz{\'a}lez}, {Burgarella}, {Dav{\'e}}, {Hirschmann}, {Somerville}, {Wuyts}, {Bromm}, {Casey}, {Fontana}, {Fujimoto}, {Gardner}, {Giavalisco}, {Grazian}, {Grogin}, {Hathi}, {Hutchison}, {Jha}, {Jogee}, {Kewley}, {Kirkpatrick}, {Long}, {Lotz}, {Pentericci}, {Pierel}, {Pirzkal}, {Ravindranath}, {Ryan}, {Trump}, {Yang}, {Bhatawdekar}, {Bisigello}, {Buat}, {Calabr{\`o}}, {Castellano}, {Cleri}, {Cooper}, {Croton}, {Daddi}, {Dekel}, {Elbaz}, {Franco}, {Gawiser}, {Holwerda}, {Huertas-Company}, {Jaskot}, {Leung}, {Lucas}, {Mobasher}, {Pandya}, {Tacchella}, {Weiner}, \& {Zavala}}]{Finkelstein23}
{Finkelstein}, S.~L., {Bagley}, M.~B., {Ferguson}, H.~C., {et~al.} 2023, \apjl, 946, L13, \dodoi{10.3847/2041-8213/acade4}

\bibitem[{{Foreman-Mackey} {et~al.}(2013){Foreman-Mackey}, {Conley}, {Meierjurgen Farr}, {Hogg}, {Lang}, {Marshall}, {Price-Whelan}, {Sanders}, \& {Zuntz}}]{Foreman13}
{Foreman-Mackey}, D., {Conley}, A., {Meierjurgen Farr}, W., {et~al.} 2013, {emcee: The MCMC Hammer}, Astrophysics Source Code Library, record ascl:1303.002.
\newblock \doeprint{1303.002}

\bibitem[{{Gehrels}(1986)}]{Gehrels86}
{Gehrels}, N. 1986, \apj, 303, 336, \dodoi{10.1086/164079}

\bibitem[{Gonzaga(2012)}]{Gonzaga.2012}
Gonzaga, S. 2012, The DrizzlePac Handbook, HST Data Handbook.
\newblock \url{http://adsabs.harvard.edu/abs/2012drzp.book.....G}

\bibitem[{{Grazian} {et~al.}(2011){Grazian}, {Castellano}, {Koekemoer}, {Fontana}, {Pentericci}, {Testa}, {Boutsia}, {Giallongo}, {Giavalisco}, \& {Santini}}]{Grazian.2011}
{Grazian}, A., {Castellano}, M., {Koekemoer}, A.~M., {et~al.} 2011, \aap, 532, A33, \dodoi{10.1051/0004-6361/201015754}

\bibitem[{Grillo {et~al.}(2015)Grillo, Suyu, Rosati, Mercurio, Balestra, Munari, Nonino, Caminha, Lombardi, Lucia, Borgani, Gobat, Biviano, Girardi, Umetsu, Coe, Koekemoer, Postman, Zitrin, Halkola, Broadhurst, Sartoris, Presotto, Annunziatella, Maier, Fritz, Vanzella, \& Frye}]{Grillo.2015}
Grillo, C., Suyu, S.~H., Rosati, P., {et~al.} 2015, ApJ, 800, 38, \dodoi{10.1088/0004-637x/800/1/38}

\bibitem[{{Harikane} {et~al.}(2024){Harikane}, {Nakajima}, {Ouchi}, {Umeda}, {Isobe}, {Ono}, {Xu}, \& {Zhang}}]{Harikane24}
{Harikane}, Y., {Nakajima}, K., {Ouchi}, M., {et~al.} 2024, \apj, 960, 56, \dodoi{10.3847/1538-4357/ad0b7e}

\bibitem[{{Harikane} {et~al.}(2022){Harikane}, {Ono}, {Ouchi}, {Liu}, {Sawicki}, {Shibuya}, {Behroozi}, {He}, {Shimasaku}, {Arnouts}, {Coupon}, {Fujimoto}, {Gwyn}, {Huang}, {Inoue}, {Kashikawa}, {Komiyama}, {Matsuoka}, \& {Willott}}]{Harikane22}
{Harikane}, Y., {Ono}, Y., {Ouchi}, M., {et~al.} 2022, \apjs, 259, 20, \dodoi{10.3847/1538-4365/ac3dfc}

\bibitem[{{Harikane} {et~al.}(2023){Harikane}, {Ouchi}, {Oguri}, {Ono}, {Nakajima}, {Isobe}, {Umeda}, {Mawatari}, \& {Zhang}}]{Harikane23}
{Harikane}, Y., {Ouchi}, M., {Oguri}, M., {et~al.} 2023, \apjs, 265, 5, \dodoi{10.3847/1538-4365/acaaa9}

\bibitem[{{Harikane} {et~al.}(2025){Harikane}, {Inoue}, {Ellis}, {Ouchi}, {Nakazato}, {Yoshida}, {Ono}, {Sun}, {Sato}, {Ferrami}, {Fujimoto}, {Kashikawa}, {McLeod}, {P{\'e}rez-Gonz{\'a}lez}, {Sawicki}, {Sugahara}, {Xu}, {Yamanaka}, {Carnall}, {Cullen}, {Dunlop}, {Egami}, {Grogin}, {Isobe}, {Koekemoer}, {Laporte}, {Lee}, {Magee}, {Matsuo}, {Matsuoka}, {Mawatari}, {Nakajima}, {Nakane}, {Tamura}, {Umeda}, \& {Yanagisawa}}]{Harikane.2025}
{Harikane}, Y., {Inoue}, A.~K., {Ellis}, R.~S., {et~al.} 2025, \apj, 980, 138, \dodoi{10.3847/1538-4357/ad9b2c}

\bibitem[{{Hathi} {et~al.}(2010){Hathi}, {Ryan}, {Cohen}, {Yan}, {Windhorst}, {McCarthy}, {O'Connell}, {Koekemoer}, {Rutkowski}, {Balick}, {Bond}, {Calzetti}, {Disney}, {Dopita}, {Frogel}, {Hall}, {Holtzman}, {Kimble}, {Paresce}, {Saha}, {Silk}, {Trauger}, {Walker}, {Whitmore}, \& {Young}}]{Hathi10}
{Hathi}, N.~P., {Ryan}, R.~E., J., {Cohen}, S.~H., {et~al.} 2010, \apj, 720, 1708, \dodoi{10.1088/0004-637X/720/2/1708}

\bibitem[{Hoag {et~al.}(2016)Hoag, Huang, Treu, Bradač, Schmidt, Wang, Brammer, Broussard, Amorin, Castellano, Fontana, Merlin, Schrabback, Trenti, \& Vulcani}]{Hoag.2016n}
Hoag, A., Huang, K.-H., Treu, T., {et~al.} 2016, The Astrophysical Journal, 831, 182, \dodoi{10.3847/0004-637X/831/2/182}

\bibitem[{{Hogg} {et~al.}(2002){Hogg}, {Baldry}, {Blanton}, \& {Eisenstein}}]{Hogg02}
{Hogg}, D.~W., {Baldry}, I.~K., {Blanton}, M.~R., \& {Eisenstein}, D.~J. 2002, arXiv e-prints, astro, \dodoi{10.48550/arXiv.astro-ph/0210394}

\bibitem[{Ishigaki {et~al.}(2015)Ishigaki, Kawamata, Ouchi, Oguri, Shimasaku, \& Ono}]{Ishigaki.2015}
Ishigaki, M., Kawamata, R., Ouchi, M., {et~al.} 2015, ApJ, 799, 12, \dodoi{10.1088/0004-637x/799/1/12}

\bibitem[{{Ishigaki} {et~al.}(2018){Ishigaki}, {Kawamata}, {Ouchi}, {Oguri}, {Shimasaku}, \& {Ono}}]{Ishigaki18}
{Ishigaki}, M., {Kawamata}, R., {Ouchi}, M., {et~al.} 2018, \apj, 854, 73, \dodoi{10.3847/1538-4357/aaa544}

\bibitem[{{Ito} {et~al.}(2020){Ito}, {Kashikawa}, {Toshikawa}, {Overzier}, {Kubo}, {Uchiyama}, {Liang}, {Onoue}, {Tanaka}, {Komiyama}, {Lee}, {Lin}, {Marinello}, {Martin}, \& {Shibuya}}]{Ito20}
{Ito}, K., {Kashikawa}, N., {Toshikawa}, J., {et~al.} 2020, \apj, 899, 5, \dodoi{10.3847/1538-4357/aba269}

\bibitem[{{Jauzac} {et~al.}(2012){Jauzac}, {Jullo}, {Kneib}, {Ebeling}, {Leauthaud}, {Ma}, {Limousin}, {Massey}, \& {Richard}}]{Jauzac.2012}
{Jauzac}, M., {Jullo}, E., {Kneib}, J.-P., {et~al.} 2012, \mnras, 426, 3369, \dodoi{10.1111/j.1365-2966.2012.21966.x}

\bibitem[{Jauzac {et~al.}(2014{\natexlab{a}})Jauzac, Clement, Limousin, Richard, Jullo, Ebeling, Atek, Kneib, Knowles, Natarajan, Eckert, Egami, Massey, \& Rexroth}]{Jauzac.2014}
Jauzac, M., Clement, B., Limousin, M., {et~al.} 2014{\natexlab{a}}, MNRAS, 443, 1549 , \dodoi{10.1093/mnras/stu1355}

\bibitem[{Jauzac {et~al.}(2014{\natexlab{b}})Jauzac, Jullo, Eckert, Ebeling, Richard, Limousin, Atek, Kneib, Clément, Egami, Harvey, Knowles, Massey, Natarajan, Neichel, \& Rexroth}]{Jauzac.2015a}
Jauzac, M., Jullo, E., Eckert, D., {et~al.} 2014{\natexlab{b}}, Monthly Notices of the Royal Astronomical Society, 446, 4132, \dodoi{10.1093/mnras/stu2425}

\bibitem[{Jauzac {et~al.}(2015)Jauzac, Richard, Jullo, Clément, Limousin, Kneib, Ebeling, Natarajan, Rodney, Atek, Massey, Eckert, Egami, \& Rexroth}]{Jauzac.2015b}
Jauzac, M., Richard, J., Jullo, E., {et~al.} 2015, Monthly Notices of the Royal Astronomical Society, 452, 1437, \dodoi{10.1093/mnras/stv1402}

\bibitem[{Johnson {et~al.}(2014)Johnson, Sharon, Bayliss, Gladders, Coe, \& Ebeling}]{Johnson.2014}
Johnson, T.~L., Sharon, K., Bayliss, M.~B., {et~al.} 2014, ApJ, 797, 48, \dodoi{10.1088/0004-637x/797/1/48}

\bibitem[{{Jullo} \& {Kneib}(2009)}]{Jullo.2009}
{Jullo}, E., \& {Kneib}, J.~P. 2009, \mnras, 395, 1319, \dodoi{10.1111/j.1365-2966.2009.14654.x}

\bibitem[{Jullo {et~al.}(2007)Jullo, Kneib, Limousin, Elíasdóttir, Marshall, \& Verdugo}]{Jullo.2007}
Jullo, E., Kneib, J.-P., Limousin, M., {et~al.} 2007, New Journal of Physics, 9, 447, \dodoi{10.1088/1367-2630/9/12/447}

\bibitem[{{Kawamata} {et~al.}(2018){Kawamata}, {Ishigaki}, {Shimasaku}, {Oguri}, {Ouchi}, \& {Tanigawa}}]{Kawamata.2018}
{Kawamata}, R., {Ishigaki}, M., {Shimasaku}, K., {et~al.} 2018, \apj, 855, 4, \dodoi{10.3847/1538-4357/aaa6cf}

\bibitem[{{Kawamata} {et~al.}(2016){Kawamata}, {Oguri}, {Ishigaki}, {Shimasaku}, \& {Ouchi}}]{Kawamata.2016}
{Kawamata}, R., {Oguri}, M., {Ishigaki}, M., {Shimasaku}, K., \& {Ouchi}, M. 2016, \apj, 819, 114, \dodoi{10.3847/0004-637X/819/2/114}

\bibitem[{Keeton(2010)}]{Keeton.2010}
Keeton, C.~R. 2010, General Relativity and Gravitation, 42, 2151 , \dodoi{10.1007/s10714-010-1041-1}

\bibitem[{{Khusanova} {et~al.}(2020){Khusanova}, {Le F{\`e}vre}, {Cassata}, {Cucciati}, {Lemaux}, {Tasca}, {Thomas}, {Garilli}, {Le Brun}, {Maccagni}, {Pentericci}, {Zamorani}, {Amor{\'\i}n}, {Bardelli}, {Castellano}, {Cassar{\`a}}, {Cimatti}, {Giavalisco}, {Hathi}, {Ilbert}, {Koekemoer}, {Marchi}, {Pforr}, {Ribeiro}, {Schaerer}, {Tresse}, {Vergani}, \& {Zucca}}]{Khusanova20}
{Khusanova}, Y., {Le F{\`e}vre}, O., {Cassata}, P., {et~al.} 2020, \aap, 634, A97, \dodoi{10.1051/0004-6361/201935400}

\bibitem[{Lagattuta {et~al.}(2017)Lagattuta, Richard, Clement, Mahler, Patricio, Pello, Soucail, Schmidt, Wisotzki, Martinez, \& Bina}]{Lagattuta.2017}
Lagattuta, D.~J., Richard, J., Clement, B., {et~al.} 2017, MNRAS, 469, 3946 , \dodoi{10.1093/mnras/stx1079}

\bibitem[{{Leethochawalit} {et~al.}(2023){Leethochawalit}, {Roberts-Borsani}, {Morishita}, {Trenti}, \& {Treu}}]{Leethochawalit23}
{Leethochawalit}, N., {Roberts-Borsani}, G., {Morishita}, T., {Trenti}, M., \& {Treu}, T. 2023, \mnras, 524, 5454, \dodoi{10.1093/mnras/stad2202}

\bibitem[{{Leethochawalit} {et~al.}(2022){Leethochawalit}, {Trenti}, {Morishita}, {Roberts-Borsani}, \& {Treu}}]{Leethochawalit.2022n}
{Leethochawalit}, N., {Trenti}, M., {Morishita}, T., {Roberts-Borsani}, G., \& {Treu}, T. 2022, \mnras, 509, 5836, \dodoi{10.1093/mnras/stab3265}

\bibitem[{{Leung} {et~al.}(2023){Leung}, {Bagley}, {Finkelstein}, {Ferguson}, {Koekemoer}, {P{\'e}rez-Gonz{\'a}lez}, {Morales}, {Kocevski}, {Yang}, {Somerville}, {Wilkins}, {Yung}, {Fujimoto}, {Larson}, {Papovich}, {Pirzkal}, {Berg}, {Lotz}, {Castellano}, {Ch{\'a}vez Ortiz}, {Cheng}, {Dickinson}, {Giavalisco}, {Hathi}, {Hutchison}, {Jung}, {Kartaltepe}, {Natarajan}, \& {Rothberg}}]{Leung23}
{Leung}, G. C.~K., {Bagley}, M.~B., {Finkelstein}, S.~L., {et~al.} 2023, \apjl, 954, L46, \dodoi{10.3847/2041-8213/acf365}

\bibitem[{{Liesenborgs} {et~al.}(2006){Liesenborgs}, {De Rijcke}, \& {Dejonghe}}]{Liesenborgs.2006}
{Liesenborgs}, J., {De Rijcke}, S., \& {Dejonghe}, H. 2006, \mnras, 367, 1209, \dodoi{10.1111/j.1365-2966.2006.10040.x}

\bibitem[{{Livermore} {et~al.}(2017){Livermore}, {Finkelstein}, \& {Lotz}}]{Livermore.2017}
{Livermore}, R.~C., {Finkelstein}, S.~L., \& {Lotz}, J.~M. 2017, \apj, 835, 113, \dodoi{10.3847/1538-4357/835/2/113}

\bibitem[{{Lotz} {et~al.}(2017){Lotz}, {Koekemoer}, {Coe}, {Grogin}, {Capak}, {Mack}, {Anderson}, {Avila}, {Barker}, {Borncamp}, {Brammer}, {Durbin}, {Gunning}, {Hilbert}, {Jenkner}, {Khandrika}, {Levay}, {Lucas}, {MacKenty}, {Ogaz}, {Porterfield}, {Reid}, {Robberto}, {Royle}, {Smith}, {Storrie-Lombardi}, {Sunnquist}, {Surace}, {Taylor}, {Williams}, {Bullock}, {Dickinson}, {Finkelstein}, {Natarajan}, {Richard}, {Robertson}, {Tumlinson}, {Zitrin}, {Flanagan}, {Sembach}, {Soifer}, \& {Mountain}}]{Lotz.2017}
{Lotz}, J.~M., {Koekemoer}, A., {Coe}, D., {et~al.} 2017, \apj, 837, 97, \dodoi{10.3847/1538-4357/837/1/97}

\bibitem[{{Mackenty} \& {Smith}(2012)}]{Mackenty.2012}
{Mackenty}, J.~W., \& {Smith}, L. 2012, CTE White Paper, Tech. Rep., STScI

\bibitem[{Madau \& Dickinson(2014)}]{Madau.2014}
Madau, P., \& Dickinson, M.~E. 2014, ARA\&A, 52, 415 , \dodoi{10.1146/annurev-astro-081811-125615}

\bibitem[{Mahler {et~al.}(2018)Mahler, Richard, Clement, Lagattuta, Schmidt, Patricio, Soucail, Bacon, Pello, Bouwens, Maseda, Martinez, Carollo, Inami, Leclercq, \& Wisotzki}]{Mahler.2018}
Mahler, G., Richard, J., Clement, B., {et~al.} 2018, MNRAS, 473, 663 , \dodoi{10.1093/mnras/stx1971}

\bibitem[{{Mascia} {et~al.}(2024){Mascia}, {Roberts-Borsani}, {Treu}, {Pentericci}, {Chen}, {Calabr{\`o}}, {Merlin}, {Paris}, {Santini}, {Brammer}, {Henry}, {Kelly}, {Mason}, {Morishita}, {Nanayakkara}, {Roy}, {Wang}, {Williams}, {Boyett}, {Brada{\v{c}}}, {Castellano}, {Glazebrook}, {Jones}, {Napolitano}, {Vulcani}, {Watson}, \& {Yang}}]{Mascia.2024}
{Mascia}, S., {Roberts-Borsani}, G., {Treu}, T., {et~al.} 2024, \aap, 690, A2, \dodoi{10.1051/0004-6361/202450493}

\bibitem[{McCully {et~al.}(2014)McCully, Keeton, Wong, \& Zabludoff}]{McCully.2014}
McCully, C., Keeton, C.~R., Wong, K.~C., \& Zabludoff, A.~I. 2014, MNRAS, 443, 3631 , \dodoi{10.1093/mnras/stu1316}

\bibitem[{{McLeod} {et~al.}(2024){McLeod}, {Donnan}, {McLure}, {Dunlop}, {Magee}, {Begley}, {Carnall}, {Cullen}, {Ellis}, {Hamadouche}, \& {Stanton}}]{Mcleod.2024}
{McLeod}, D.~J., {Donnan}, C.~T., {McLure}, R.~J., {et~al.} 2024, \mnras, 527, 5004, \dodoi{10.1093/mnras/stad3471}

\bibitem[{Mehta {et~al.}(2017)Mehta, Scarlata, Rafelski, Gburek, Teplitz, Alavi, Boylan-Kolchin, Finkelstein, Gardner, Grogin, Koekemoer, Kurczynski, Siana, Codoreanu, Mello, Lee, \& Soto}]{Mehta.2017}
Mehta, V., Scarlata, C., Rafelski, M.~A., {et~al.} 2017, ApJ, 838, 29, \dodoi{10.3847/1538-4357/aa6259}

\bibitem[{{Merten} {et~al.}(2009){Merten}, {Cacciato}, {Meneghetti}, {Mignone}, \& {Bartelmann}}]{Merten.2009}
{Merten}, J., {Cacciato}, M., {Meneghetti}, M., {Mignone}, C., \& {Bartelmann}, M. 2009, \aap, 500, 681, \dodoi{10.1051/0004-6361/200810372}

\bibitem[{Merten {et~al.}(2011)Merten, Coe, Dupke, Massey, Zitrin, Cypriano, Okabe, Frye, Braglia, Jimenez-Teja, Benitez, Broadhurst, Rhodes, Meneghetti, Moustakas, Sodré, Krick, \& Bregman}]{Merten.2011}
Merten, J.~C., Coe, D.~A., Dupke, R., {et~al.} 2011, MNRAS, 417, 333 , \dodoi{10.1111/j.1365-2966.2011.19266.x}

\bibitem[{{Mohammed} {et~al.}(2014){Mohammed}, {Liesenborgs}, {Saha}, \& {Williams}}]{Mohammed.2014}
{Mohammed}, I., {Liesenborgs}, J., {Saha}, P., \& {Williams}, L. L.~R. 2014, \mnras, 439, 2651, \dodoi{10.1093/mnras/stu124}

\bibitem[{{Moutard} {et~al.}(2020){Moutard}, {Sawicki}, {Arnouts}, {Golob}, {Coupon}, {Ilbert}, {Yang}, \& {Gwyn}}]{Moutard20}
{Moutard}, T., {Sawicki}, M., {Arnouts}, S., {et~al.} 2020, \mnras, 494, 1894, \dodoi{10.1093/mnras/staa706}

\bibitem[{{Oesch} {et~al.}(2010){Oesch}, {Bouwens}, {Carollo}, {Illingworth}, {Magee}, {Trenti}, {Stiavelli}, {Franx}, {Labb{\'e}}, \& {van Dokkum}}]{Oesch10}
{Oesch}, P.~A., {Bouwens}, R.~J., {Carollo}, C.~M., {et~al.} 2010, \apjl, 725, L150, \dodoi{10.1088/2041-8205/725/2/L150}

\bibitem[{Oguri(2010)}]{Oguri.2010}
Oguri, M. 2010, Publications of the Astronomical Society of Japan, 62, 1017 , \dodoi{10.1093/pasj/62.4.1017}

\bibitem[{{Oke} \& {Gunn}(1983)}]{Oke83}
{Oke}, J.~B., \& {Gunn}, J.~E. 1983, \apj, 266, 713, \dodoi{10.1086/160817}

\bibitem[{Owers {et~al.}(2011)Owers, Randall, Nulsen, Couch, David, \& Kempner}]{Owers.2011}
Owers, M.~S., Randall, S.~W., Nulsen, P. E.~J., {et~al.} 2011, ApJ, 728, 27, \dodoi{10.1088/0004-637x/728/1/27}

\bibitem[{{Page} {et~al.}(2021){Page}, {Dwelly}, {McHardy}, {Seymour}, {Mason}, {Sharma}, {Kennea}, {Sasseen}, {Rawlings}, {Breeveld}, {Ferreras}, {Loaring}, {Walton}, \& {Symeonidis}}]{Page21}
{Page}, M.~J., {Dwelly}, T., {McHardy}, I., {et~al.} 2021, \mnras, 506, 473, \dodoi{10.1093/mnras/stab1638}

\bibitem[{{Page} {et~al.}(2025){Page}, {Dwelly}, {McHardy}, {Seymour}, {Mason}, {Sharma}, {Kennea}, {Sasseen}, {Breeveld}, \& {Matthews}}]{Page.2025}
---. 2025, \mnras, 536, 518, \dodoi{10.1093/mnras/stae2498}

\bibitem[{Pagul {et~al.}(2021)Pagul, Sánchez, Davidzon, \& Mobasher}]{Pagul.2021}
Pagul, A., Sánchez, F.~J., Davidzon, I., \& Mobasher, B. 2021, ApJS, 256, 27, \dodoi{10.3847/1538-4365/abea9d}

\bibitem[{{P{\'e}rez-Gonz{\'a}lez} {et~al.}(2023){P{\'e}rez-Gonz{\'a}lez}, {Costantin}, {Langeroodi}, {Rinaldi}, {Annunziatella}, {Ilbert}, {Colina}, {N{\o}rgaard-Nielsen}, {Greve}, {{\"O}stlin}, {Wright}, {Alonso-Herrero}, {{\'A}lvarez-M{\'a}rquez}, {Caputi}, {Eckart}, {Le F{\`e}vre}, {Labiano}, {Garc{\'\i}a-Mar{\'\i}n}, {Hjorth}, {Kendrew}, {Pye}, {Tikkanen}, {van der Werf}, {Walter}, {Ward}, {Bik}, {Boogaard}, {Bosman}, {G{\'o}mez}, {Gillman}, {Iani}, {Jermann}, {Melinder}, {Meyer}, {Moutard}, {van Dishoek}, {Henning}, {Lagage}, {Guedel}, {Peissker}, {Ray}, {Vandenbussche}, {Garc{\'\i}a-Argum{\'a}nez}, \& {Mar{\'\i}a M{\'e}rida}}]{Perez23}
{P{\'e}rez-Gonz{\'a}lez}, P.~G., {Costantin}, L., {Langeroodi}, D., {et~al.} 2023, \apjl, 951, L1, \dodoi{10.3847/2041-8213/acd9d0}

\bibitem[{{P{\'e}rez-Gonz{\'a}lez} {et~al.}(2025){P{\'e}rez-Gonz{\'a}lez}, {{\"O}stlin}, {Costantin}, {Melinder}, {Finkelstein}, {Somerville}, {Annunziatella}, {{\'A}lvarez-M{\'a}rquez}, {Colina}, {Dekel}, {Ferguson}, {Li}, {Yung}, {Bagley}, {Boogaard}, {Burgarella}, {Calabr{\`o}}, {Caputi}, {Cheng}, {Dickinson}, {Eckart}, {Giavalisco}, {Gillman}, {Greve}, {Hamed}, {Hathi}, {Hjorth}, {Huertas-Company}, {Kartaltepe}, {Koekemoer}, {Kokorev}, {Labiano}, {Langeroodi}, {Leung}, {Natarajan}, {Papovich}, {Peissker}, {Pentericci}, {Pirzkal}, {Rinaldi}, {van der Werf}, \& {Walter}}]{Perez.2025}
{P{\'e}rez-Gonz{\'a}lez}, P.~G., {{\"O}stlin}, G., {Costantin}, L., {et~al.} 2025, \apj, 991, 179, \dodoi{10.3847/1538-4357/adf8c9}

\bibitem[{Postman {et~al.}(2012)Postman, Coe, Benítez, Bradley, Broadhurst, Donahue, Ford, Graur, Graves, Jouvel, Koekemoer, Lemze, Medezinski, Molino, Moustakas, Ogaz, Riess, Rodney, Rosati, Umetsu, Zheng, Zitrin, Bartelmann, Bouwens, Czakon, Golwala, Host, Infante, Jha, Jimenez-Teja, Kelson, Lahav, Lazkoz, Maoz, McCully, Melchior, Meneghetti, Merten, Moustakas, Nonino, Patel, Regös, Sayers, Seitz, \& Wel}]{Postman.2012}
Postman, M., Coe, D.~A., Benítez, N., {et~al.} 2012, ApJS, 199, 25, \dodoi{10.1088/0067-0049/199/2/25}

\bibitem[{Prichard {et~al.}(2022)Prichard, Rafelski, Cooke, Meštrić, Bassett, Ryan-Weber, Sunnquist, Alavi, Hathi, Wang, Revalski, Bajaj, O’Meara, \& Spitler}]{Prichard.2022}
Prichard, L.~J., Rafelski, M., Cooke, J., {et~al.} 2022, The Astrophysical Journal, 924, 14, \dodoi{10.3847/1538-4357/ac3004}

\bibitem[{Rafelski {et~al.}(2015)Rafelski, Teplitz, Gardner, Coe, Bond, Koekemoer, Grogin, Kurczynski, McGrath, Bourque, Atek, Brown, Colbert, Codoreanu, Ferguson, Finkelstein, Gawiser, Giavalisco, Gronwall, Hanish, Lee, Mehta, Mello, Ravindranath, Ryan, Scarlata, Siana, Soto, \& Voyer}]{Rafelski.2015}
Rafelski, M.~A., Teplitz, H.~I., Gardner, J.~P., {et~al.} 2015, AJ, 150, 31, \dodoi{10.1088/0004-6256/150/1/31}

\bibitem[{{Rees} \& {Ostriker}(1977)}]{Rees77}
{Rees}, M.~J., \& {Ostriker}, J.~P. 1977, \mnras, 179, 541, \dodoi{10.1093/mnras/179.4.541}

\bibitem[{Richard {et~al.}(2014)Richard, Jauzac, Limousin, Jullo, Clement, Ebeling, Kneib, Atek, Natarajan, Egami, Livermore, \& Bower}]{Richard.2014}
Richard, J., Jauzac, M., Limousin, M., {et~al.} 2014, MNRAS, 444, 268 , \dodoi{10.1093/mnras/stu1395}

\bibitem[{{Robertson} {et~al.}(2024){Robertson}, {Johnson}, {Tacchella}, {Eisenstein}, {Hainline}, {Arribas}, {Baker}, {Bunker}, {Carniani}, {Cargile}, {Carreira}, {Charlot}, {Chevallard}, {Curti}, {Curtis-Lake}, {D'Eugenio}, {Egami}, {Hausen}, {Helton}, {Jakobsen}, {Ji}, {Jones}, {Maiolino}, {Maseda}, {Nelson}, {P{\'e}rez-Gonz{\'a}lez}, {Pusk{\'a}s}, {Rieke}, {Smit}, {Sun}, {{\"U}bler}, {Whitler}, {Williams}, {Willmer}, {Willott}, \& {Witstok}}]{Robertson.2024}
{Robertson}, B., {Johnson}, B.~D., {Tacchella}, S., {et~al.} 2024, \apj, 970, 31, \dodoi{10.3847/1538-4357/ad463d}

\bibitem[{{Rojas-Ruiz} {et~al.}(2020){Rojas-Ruiz}, {Finkelstein}, {Bagley}, {Stevans}, {Finkelstein}, {Larson}, {Mechtley}, \& {Diekmann}}]{Rojas20}
{Rojas-Ruiz}, S., {Finkelstein}, S.~L., {Bagley}, M.~B., {et~al.} 2020, \apj, 891, 146, \dodoi{10.3847/1538-4357/ab7659}

\bibitem[{{Schechter}(1976)}]{Schechter76}
{Schechter}, P. 1976, \apj, 203, 297, \dodoi{10.1086/154079}

\bibitem[{Sharma {et~al.}(2022)Sharma, Page, \& Breeveld}]{Sharma.2022}
Sharma, M., Page, M.~J., \& Breeveld, A.~A. 2022, Monthly Notices of the Royal Astronomical Society, 511, 4882, \dodoi{10.1093/mnras/stac356}

\bibitem[{Sharma {et~al.}(2024)Sharma, Page, Ferreras, \& Breeveld}]{Sharma.2024}
Sharma, M., Page, M.~J., Ferreras, I., \& Breeveld, A.~A. 2024, Monthly Notices of the Royal Astronomical Society, 531, 2040, \dodoi{10.1093/mnras/stae1278}

\bibitem[{{Sheth} \& {Tormen}(1999)}]{Sheth99}
{Sheth}, R.~K., \& {Tormen}, G. 1999, \mnras, 308, 119, \dodoi{10.1046/j.1365-8711.1999.02692.x}

\bibitem[{Shibuya {et~al.}(2015)Shibuya, Ouchi, \& Harikane}]{Shibuya.2015}
Shibuya, T., Ouchi, M., \& Harikane, Y. 2015, ApJS, 219, 15, \dodoi{10.1088/0067-0049/219/2/15}

\bibitem[{{Soucail} {et~al.}(1987){Soucail}, {Fort}, {Mellier}, \& {Picat}}]{Soucail.1987}
{Soucail}, G., {Fort}, B., {Mellier}, Y., \& {Picat}, J.~P. 1987, \aap, 172, L14

\bibitem[{Sun {et~al.}(2024)Sun, Wang, Teplitz, Mehta, Alavi, Rafelski, Windhorst, Scarlata, Gardner, Smith, Sunnquist, Prichard, Cheng, Grogin, Hathi, Hayes, Koekemoer, Mobasher, Nedkova, O’Connell, Robertson, Taamoli, Yung, Brammer, Colbert, Conselice, Gawiser, Guo, Jansen, Ji, Lucas, Rutkowski, Siana, Vanzella, Ashcraft, Bagley, Baronchelli, Barro, Blanche, Broussard, Carleton, Chartab, Codoreanu, Cohen, Dai, Darvish, Davé, DeGroot, De~Mello, Dickinson, Emami, Ferguson, Ferreira, Finkelstein, Finkelstein, Gburek, Giavalisco, Grazian, Gronwall, Hemmati, Howell, Iyer, Kaviraj, Kurczynski, Lazar, MacKenty, Mantha, Martin, Martin, McCabe, Olsen, Otteson, Ravindranath, Redshaw, Sattari, Soto, Zabelle, \& the UVCANDELS~team}]{Sun.2024}
Sun, L., Wang, X., Teplitz, H.~I., {et~al.} 2024, The Astrophysical Journal, 972, 8, \dodoi{10.3847/1538-4357/ad5540}

\bibitem[{Teplitz {et~al.}(2013)Teplitz, Rafelski, Kurczynski, Bond, Grogin, Koekemoer, Atek, Brown, Coe, Colbert, Ferguson, Finkelstein, Gardner, Gawiser, Giavalisco, Gronwall, Hanish, Lee, Mello, Ravindranath, Ryan, Siana, Scarlata, Soto, Voyer, \& Wolfe}]{Teplitz.2013}
Teplitz, H.~I., Rafelski, M.~A., Kurczynski, P., {et~al.} 2013, AJ, 146, 159, \dodoi{10.1088/0004-6256/146/6/159}

\bibitem[{{Trenti} \& {Stiavelli}(2008)}]{Trenti08}
{Trenti}, M., \& {Stiavelli}, M. 2008, \apj, 676, 767, \dodoi{10.1086/528674}

\bibitem[{{Varadaraj} {et~al.}(2023){Varadaraj}, {Bowler}, {Jarvis}, {Adams}, \& {H{\"a}u{\ss}ler}}]{Varadaraj23}
{Varadaraj}, R.~G., {Bowler}, R.~A.~A., {Jarvis}, M.~J., {Adams}, N.~J., \& {H{\"a}u{\ss}ler}, B. 2023, \mnras, 524, 4586, \dodoi{10.1093/mnras/stad2081}

\bibitem[{Wang {et~al.}(2020)Wang, Jones, Treu, Daddi, Brammer, Sharon, Morishita, Abramson, Colbert, Henry, Hopkins, Malkan, Schmidt, Teplitz, \& Vulcani}]{Wang.2020}
Wang, X., Jones, T.~A., Treu, T.~L., {et~al.} 2020, ApJ, 900, 183, \dodoi{10.3847/1538-4357/abacce}

\bibitem[{{Watson} {et~al.}(2025){Watson}, {Vulcani}, {Treu}, {Roberts-Borsani}, {Dalmasso}, {He}, {Malkan}, {Morishita}, {Rojas Ruiz}, {Zhang}, {Acharyya}, {Bergamini}, {Brada{\v{c}}}, {Fontana}, {Grillo}, {Jones}, {Marchesini}, {Nanayakkara}, {Pentericci}, {Tubthong}, \& {Wang}}]{Watson.2025}
{Watson}, P.~J., {Vulcani}, B., {Treu}, T., {et~al.} 2025, \aap, 699, A225, \dodoi{10.1051/0004-6361/202554954}

\bibitem[{{Weisz} {et~al.}(2014){Weisz}, {Johnson}, \& {Conroy}}]{Weisz14}
{Weisz}, D.~R., {Johnson}, B.~D., \& {Conroy}, C. 2014, \apjl, 794, L3, \dodoi{10.1088/2041-8205/794/1/L3}

\bibitem[{{White} \& {Rees}(1978)}]{White78}
{White}, S.~D.~M., \& {Rees}, M.~J. 1978, \mnras, 183, 341, \dodoi{10.1093/mnras/183.3.341}

\bibitem[{{Whitler} {et~al.}(2025){Whitler}, {Stark}, {Topping}, {Robertson}, {Rieke}, {Hainline}, {Endsley}, {Chen}, {Baker}, {Bhatawdekar}, {Bunker}, {Carniani}, {Charlot}, {Chevallard}, {Curtis-Lake}, {Egami}, {Eisenstein}, {Helton}, {Ji}, {Johnson}, {P{\'e}rez-Gonz{\'a}lez}, {Rinaldi}, {Tacchella}, {Williams}, {Willmer}, {Willott}, \& {Witstok}}]{Whitler.2025}
{Whitler}, L., {Stark}, D.~P., {Topping}, M.~W., {et~al.} 2025, \apj, 992, 63, \dodoi{10.3847/1538-4357/adfddc}

\bibitem[{Williams {et~al.}(2018)Williams, Curtis-Lake, Hainline, Chevallard, Robertson, Charlot, Endsley, Stark, Willmer, Alberts, Amorin, Arribas, Baum, Bunker, Carniani, Crandall, Egami, Eisenstein, Ferruit, Husemann, Maseda, Maiolino, Rawle, Rieke, Smit, Tacchella, \& Willott}]{Williams18}
Williams, C.~C., Curtis-Lake, E., Hainline, K.~N., {et~al.} 2018, The Astrophysical Journal Supplement Series, 236, 33, \dodoi{10.3847/1538-4365/aabcbb}

\bibitem[{{Willott} {et~al.}(2024){Willott}, {Desprez}, {Asada}, {Sarrouh}, {Abraham}, {Brada{\v{c}}}, {Brammer}, {Estrada-Carpenter}, {Iyer}, {Martis}, {Matharu}, {Mowla}, {Muzzin}, {Noirot}, {Sawicki}, {Strait}, {Rihtar{\v{s}}i{\v{c}}}, \& {Withers}}]{Willott.2024}
{Willott}, C.~J., {Desprez}, G., {Asada}, Y., {et~al.} 2024, \apj, 966, 74, \dodoi{10.3847/1538-4357/ad35bc}

\bibitem[{{Yue} {et~al.}(2018){Yue}, {Castellano}, {Ferrara}, {Fontana}, {Merlin}, {Amor{\'\i}n}, {Grazian}, {M{\'a}rmol-Queralto}, {Micha{\l}owski}, {Mortlock}, {Paris}, {Parsa}, {Pilo}, {Santini}, \& {Di Criscienzo}}]{Yue.2018n}
{Yue}, B., {Castellano}, M., {Ferrara}, A., {et~al.} 2018, \apj, 868, 115, \dodoi{10.3847/1538-4357/aae77f}

\bibitem[{{Yung} {et~al.}(2020{\natexlab{a}}){Yung}, {Somerville}, {Finkelstein}, {Popping}, {Dav{\'e}}, {Venkatesan}, {Behroozi}, \& {Ferguson}}]{Yung20b}
{Yung}, L.~Y.~A., {Somerville}, R.~S., {Finkelstein}, S.~L., {et~al.} 2020{\natexlab{a}}, \mnras, 496, 4574, \dodoi{10.1093/mnras/staa1800}

\bibitem[{{Yung} {et~al.}(2020{\natexlab{b}}){Yung}, {Somerville}, {Popping}, \& {Finkelstein}}]{Yung20a}
{Yung}, L.~Y.~A., {Somerville}, R.~S., {Popping}, G., \& {Finkelstein}, S.~L. 2020{\natexlab{b}}, \mnras, 494, 1002, \dodoi{10.1093/mnras/staa714}

\bibitem[{{Zhang} {et~al.}(2021){Zhang}, {Ouchi}, {Gebhardt}, {Mentuch Cooper}, {Liu}, {Davis}, {Jeong}, {Farrow}, {Finkelstein}, {Gawiser}, {Hill}, {Harikane}, {Kakuma}, {Acquaviva}, {Casey}, {Fabricius}, {Hopp}, {Jarvis}, {Landriau}, {Mawatari}, {Mukae}, {Ono}, {Sakai}, \& {Schneider}}]{Zhang21}
{Zhang}, Y., {Ouchi}, M., {Gebhardt}, K., {et~al.} 2021, \apj, 922, 167, \dodoi{10.3847/1538-4357/ac1e97}

\bibitem[{Zitrin {et~al.}(2009)Zitrin, Broadhurst, Umetsu, Coe, Benítez, Ascaso, Bradley, Ford, Jee, Medezinski, Rephaeli, \& Zheng}]{Zitrin.2009}
Zitrin, A., Broadhurst, T., Umetsu, K., {et~al.} 2009, MNRAS, 396, 1985 , \dodoi{10.1111/j.1365-2966.2009.14899.x}

\bibitem[{Zitrin {et~al.}(2013)Zitrin, Meneghetti, Umetsu, Broadhurst, Bartelmann, Bouwens, Bradley, Carrasco, Coe, Ford, Kelson, Koekemoer, Medezinski, Moustakas, Moustakas, Nonino, Postman, Rosati, Seidel, Seitz, Sendra, Shu, Vega, \& Zheng}]{Zitrin.2013}
Zitrin, A., Meneghetti, M., Umetsu, K., {et~al.} 2013, ApJ, 762, L30, \dodoi{10.1088/2041-8205/762/2/l30}

\end{thebibliography}

\end{document}